\documentclass[12pt]{article}

\usepackage{cite}
\usepackage{axodraw}
\usepackage[dvips]{graphicx}

\def\theequation{\arabic{section}.\arabic{equation}}

\begin{document}

\begin{flushright}
MAN/HEP/2008/09\\[-2pt]
{\tt arXiv:0805.1677}\\
May 2008
\end{flushright}
\bigskip

\begin{center}
{\LARGE {\bf Electroweak Resonant Leptogenesis}}\\[0.3cm]
{\LARGE {\bf in the Singlet Majoron Model}}\\[1.5cm]
{\large Apostolos Pilaftsis}\\[0.3cm] 
{\em School of Physics and Astronomy, University of Manchester,}\\ 
{\em Manchester M13 9PL, United Kingdom}
\end{center}

\vspace{1.5cm} 

\centerline{\bf ABSTRACT}

\noindent
We study resonant leptogenesis  at the electroweak phase transition in
the singlet Majoron model  with right-handed neutrinos.  We consider a
scenario, where  the SM gauge group  and the lepton  number break down
spontaneously  during  a  second-order electroweak  phase  transition.
We~calculate   the   flavour-   and   temperature-dependent   leptonic
asymmetries, by including the  novel contributions from the transverse
polarisations of the $W^\pm$ and $Z$~bosons.  The~required resummation
of  the  gauge-dependent  off-shell  heavy-neutrino  self-energies  is
consistently treated within the gauge-invariant framework of the Pinch
Technique.   Taking  into  consideration  the freeze-out  dynamics  of
sphalerons,  we delineate  the parameter  space of  the model  that is
compatible  with successful  electroweak  resonant leptogenesis.   The
phenomenological  and  astrophysical  implications  of the  model  are
discussed.


\newpage

\setcounter{equation}{0}
\section{Introduction}

Leptogenesis~\cite{FY} provides  an elegant framework  to consistently
address    the   observed   Baryon    Asymmetry   in    the   Universe
(BAU)~\cite{WMAP}    in   minimal    extensions   of    the   Standard
Model~(SM)~\cite{reviews}.   According  to  the standard  paradigm  of
leptogenesis, there exist heavy  Majorana neutrinos of masses close to
the Grand Unified  Theory (GUT) scale $M_{\rm GUT}  \sim 10^{16}$ that
decay  out of equilibrium  and create  a net  excess of  lepton number
$(L)$, which  gets reprocessed into the observed  baryon number $(B)$,
through the  $(B+L)$-violating sphaleron interactions~\cite{KRS}.  The
attractive  feature of  such a  scenario is  that the  GUT-scale heavy
Majorana neutrinos  could also explain the observed  smallness in mass
of  the  SM   light  neutrinos  by  means  of   the  so-called  seesaw
mechanism~\cite{seesaw}.

The  original  GUT-scale  leptogenesis  scenario, however,  runs  into
certain difficulties, when one attempts to explain the flatness of the
Universe and  other cosmological data~\cite{WMAP}  within supergravity
models   of  inflation.    To  avoid   overproduction   of  gravitinos
$\widetilde{G}$ whose late decays  may ruin the successful predictions
of  Big Bang  Nucleosynthesis  (BBN), the  reheat temperature  $T_{\rm
reh}$  of the Universe  should be  lower than  $10^9$--$10^6$~GeV, for
$m_{\widetilde{G}} =  8$--0.2~TeV~\cite{gravitino}.  This implies that
the heavy Majorana neutrinos should  accordingly have masses as low as
$T_{\rm  reh} \stackrel{<}{{}_\sim}  10^9$~GeV, thereby  rendering the
relation of  these particles with GUT-scale physics  less natural.  On
the  other  hand, it  proves  very  difficult  to directly  probe  the
heavy-neutrino  sector  of  such  a model  at  high-energy  colliders,
e.g.~at the LHC or ILC, or in any other foreseeable experiment.

A  potentially  interesting solution  to  the  above  problems may  be
obtained    within   the    framework    of   resonant    leptogenesis
(RL)~\cite{APRD}.  The  key aspect of  RL is that  self-energy effects
dominate  the  leptonic  asymmetries~\cite{LiuSegre}, when  two  heavy
Majorana neutrinos happen to have a small mass difference with respect
to their actual masses.  If this mass difference becomes comparable to
the  heavy neutrino  widths, a  resonant enhancement  of  the leptonic
asymmetries    takes   place    that   may    reach    values   ${\cal
O}(1)$~\cite{APRD,PU}.  An indispensable feature  of RL models is that
flavour   effects   due   to   the  light-to-heavy   neutrino   Yukawa
couplings~\cite{EMX}  play   a  dramatic  role  and   can  modify  the
predictions for the BAU  by many orders of magnitude~\cite{APtau,PU2}.
Most    importantly,     these    flavour    effects     enable    the
modelling~\cite{PU2}  of minimal  RL scenarios  with electroweak-scale
heavy   Majorana    neutrinos   that   could   be    tested   at   the
LHC~\cite{APZPC,DGP} and  in other non-accelerator  experiments, while
maintaining  agreement  with   the  low-energy  neutrino  data.   Many
variants     of      RL     have     been      proposed     in     the
literature~\cite{RLpapers,RLextra},           including           soft
leptogenesis~\cite{soft} and radiative leptogenesis~\cite{rad}.

In spite  of the  many existing studies,  leptogenesis models  face in
general a serious restriction concerning the origin of the required CP
and $L$ violation.  If CP or $L$ violation were due to the spontaneous
symmetry breaking  (SSB) of  the SM gauge  group, a net  $L$ asymmetry
could  only  be  generated  during the  electroweak  phase  transition
(EWPT), provided the  heavy Majorana neutrinos are not  too heavy such
that  they  have not  already  decayed  away  while the  Universe  was
expanding.\footnote{An exception  to this  argument may result  from a
phase  transition  that  is  strongly  first order.  However,  such  a
scenario   is    not   feasible    within   the   SM    with   singlet
neutrinos~\cite{HR} (see also our discussion below).}

In this paper we show how RL constitutes an interesting alternative to
provide  a  viable  solution  to  the  above  problem  as  well.   For
definiteness,  we  consider  a   minimal  extension  of  the  SM  with
right-handed  neutrinos and  a  complex singlet  field $\Sigma$.   The
model  possesses   a  global  lepton  symmetry   U(1)$_l$  which  gets
spontaneously  broken through  the vacuum  expectation value  (VEV) of
$\Sigma$, giving  rise to  the usual $\Delta  L = 2$  Majorana masses.
Because of the SSB of the U(1)$_l$, the model predicts a true massless
Goldstone boson,  the Majoron. Therefore, this scenario  is called the
singlet Majoron  model in the  literature~\cite{CMP,APMaj}.  Depending
on  the  particular structure  of  the  Higgs  potential, the  VEV  of
$\Sigma$ may  be related to  the VEV of  the SM Higgs  doublet $\Phi$.
Such  a  relation,  for  example,  arises  if  the  bilinear  operator
$\Sigma^*\Sigma$ is small or absent from the Higgs potential.  In this
case, the breaking of $L$ occurs during the EWPT.  For the model under
study and  given the LEP  limit~\cite{LEPHiggs} on the SM  Higgs boson
$M_H \stackrel{>}{{}_\sim} 115$~GeV, the EWPT is expected to be second
order  and hence  continuous from  the symmetric  phase to  the broken
one~\cite{EWSM}.

We should now  notice that all SM fermions  and right-handed neutrinos
have no chiral masses above the EWPT and therefore the generation of a
net leptonic  asymmetry is not possible.  Consequently,  in this model
successful baryogenesis can result from  RL at the EWPT.  Although the
singlet  Majoron model  that  we  will be  studying  here violates  CP
explicitly, the results of our analysis can straightforwardly apply to
models  with an  extended  Higgs sector  that  realise spontaneous  CP
violation at the electroweak scale.

The  paper is organised  as follows:  Section~\ref{sec:model} presents
the  basic features  of the  singlet Majoron  model  with right-handed
neutrinos, including the interaction  Lagrangians that are relevant to
the calculation of the leptonic asymmetries in Section~\ref{sec:asym}.
Moreover,   in   Section~\ref{sec:asym}    we   consider   the   novel
contributions to the leptonic  asymmetries, coming from the transverse
polarisations of the $W^\pm$ and  $Z$ bosons. In the same context, the
resummation   of    the   gauge-dependent   off-shell   heavy-neutrino
self-energies~\cite{PP,APNPB}  (which remains  an  essential operation
in~RL)  is  performed  within   the  so-called  Pinch  Technique  (PT)
framework~\cite{PT}.    In  Section~\ref{sec:EWPT}   we   analyse  the
Boltzmann  dynamics  of  the  sphaleron  effects  on  RL  and  present
predictions  for the  BAU.  Section~\ref{sec:pheno}  is devoted  to the
phenomenological and astrophysical implications of the singlet Majoron
model.  Finally, Section~\ref{sec:concl} contains our conclusions.

\setcounter{equation}{0}
\section{The Singlet Majoron Model}\label{sec:model}

Here  we   describe  the  basic   features  of  the   singlet  Majoron
model~\cite{CMP,APMaj} augmented  with a number  n$_R$ of right-handed
neutrinos $\nu_{\alpha  R}$ (with  $\alpha = 1,2,  \dots, \mbox{n}_R$)
that will be relevant to our study.  As mentioned in the introduction,
the singlet Majoron model  contains one complex singlet field $\Sigma$
in addition to the SM  Higgs doublet $\Phi$.  Although $\Sigma$ is not
charged  under  the SM  gauge  group SU(2)$_L\,\otimes$\,U(1)$_Y$,  it
still  carries  a non-zero  quantum  number  under  the global  lepton
symmetry U(1)$_l$.  More explicitly, the scalar potential of the model
is given by
\begin{eqnarray}
  \label{LV}
-\, {\cal L}_V \!&=&\! m^2_\Phi\, \Phi^\dagger \Phi\ +\ m^2_\Sigma\, 
\Sigma^* \Sigma\ +\ \frac{\lambda_\Phi}{2}\, (\Phi^\dagger \Phi)^2 \ +\ 
\frac{\lambda_\Sigma}{2}\, (\Sigma^* \Sigma)^2 \ -\
\delta\, \Phi^\dagger \Phi\, \Sigma^* \Sigma\; .\qquad
\end{eqnarray}
In  order  to minimise  the  potential~(\ref{LV}),  we first  linearly
decompose the scalar fields as follows:
\begin{equation}
  \label{PhiSigma}
\Phi\ =\ \left ( \begin{array}{c} G^+ \\ \displaystyle{\frac{v}{\sqrt{2}}\ +\ 
\frac{\phi\: +\: iG}{\sqrt{2}}} \end{array} \right)\; , \qquad 
\Sigma\ =\ \frac{w}{\sqrt{2}}\ +\ \frac{\sigma\: +\: iJ}{\sqrt{2}}\ .
\end{equation}
Then, the extremal or tadpole conditions may easily be calculated by
\begin{eqnarray}
  \label{TadH}
T_\phi \!&\equiv&\! -\; \Bigg< \frac{\partial {\cal L}_V}{\partial
  \phi}\Bigg>\ =\  v\, \Bigg(\, m^2_\Phi\: +\:
  \frac{\lambda_\Phi}{2}\, v^2\: -\: \frac{\delta}{2}\, w^2\, \Bigg)\
  =\ 0\; ,\\
  \label{TadS}
T_\sigma \!&\equiv&\! -\; \Bigg< \frac{\partial {\cal L}_V}{\partial
  \sigma}\Bigg>\ =\  w\, \Bigg(\, m^2_\Sigma\: +\:
  \frac{\lambda_\Sigma}{2}\, w^2\: -\: \frac{\delta}{2}\, v^2\,
  \Bigg)\ =\ 0\; .
\end{eqnarray}
If   $m^2_{\Phi}$   or   $m^2_\Sigma$   are  negative,   the   tadpole
conditions~(\ref{TadH}) and~(\ref{TadS})  imply that the  ground state
of   the    scalar   potential   breaks    spontaneously   the   local
SU(2)$_L\,\otimes$\,U(1)$_Y$  and   the  global  U(1)$_l$  symmetries,
through the non-zero VEVs $v$ and $w$, respectively.

Expanding the fields  $\Phi$ and $\Sigma$ about their  VEVs, we obtain
three would-be  Goldstone bosons $G^\pm$  and $G^0$, which  become the
longitudinal  polarisations of $W^\pm$  and $Z$  bosons, and  one true
massless  Goldstone boson  $J$ associated  with the  SSB  of U(1)$_l$.
This  massless  CP-odd  field  $J$   is  called  the  Majoron  in  the
literature~\cite{CMP,APMaj}.  In addition, there are two CP-even Higgs
fields $H$ and $S$, whose masses are determined by the diagonalisation
of the mass matrix
\begin{equation}
  \label{M2}
{\cal M}^2\ =\ \left( \begin{array}{cc}
\lambda_\Phi\, v^2 & -\delta\, v w \\
-\delta\, v w & \lambda_\Sigma\, w^2
\end{array}\right)\; ,
\end{equation}
where ${\cal M}^2$ is defined in the weak basis $(\phi \,,\ \sigma )$.
The  Higgs mass  eigenstates $H$  and $S$  are related  to  the states
$\phi$ and $\sigma$, through the orthogonal transformation:
\begin{equation}
 \label{Mix}
\left( \begin{array}{c} \phi \\ \sigma \end{array} \right)\ =\ 
\left( \begin{array}{cc} c_\theta & - s_\theta \\
                 s_\theta & c_\theta \end{array} \right)\ \left(
\begin{array}{c} H \\ S \end{array} \right) \ ,
\end{equation}
with $t_\beta = s_\beta/c_\beta = v/w$ and  
\begin{equation}
  \label{t2theta}
t_{2\theta}  \ =\ \frac{2\,\delta\, t_\beta}{\lambda_\Sigma\: 
-\: \lambda_\Phi t^2_\beta}\ .
\end{equation}
In the  above we  used the short-hand  notation: $s_x \equiv  \sin x$,
$c_x \equiv  \cos x$  and $t_x \equiv  \tan x$. Moreover,  the squared
mass  eigenvalues of  the CP-even  $H$ and  $S$ bosons  may  easily be
calculated from ${\cal M}^2$ in~(\ref{M2}) and are given by
\begin{equation}
  \label{MHS2}
M^2_{H,S}\ =\ \frac{v^2}{2}\, \Bigg[\, \lambda_\Phi\: +\:
\lambda_\Sigma\,t^{-2}_\beta\ \pm\ \sqrt{ \Big(\lambda_\Phi\, -\,
\lambda_\Sigma\,t^{-2}_\beta\Big)^2\: +\: \delta^2\, t^{-2}_\beta }\ \Bigg]\; .
\end{equation}
The  requirement  that  $M^2_{H,S}$  be  positive gives  rise  to  the
inequality conditions,
\begin{equation}
  \label{stable}
\lambda_{\Phi,\Sigma}\ >\ 0\; ,\qquad 
\lambda_\Phi\, \lambda_\Sigma\ >\ \delta^2\; ,
\end{equation}
for the quartic  couplings of the potential. In  this context, we note
that if $|m^2_\Sigma| \ll  (\delta / \lambda_\Phi )\, |m^2_\Phi|$ such
that $m^2_\Sigma$ can be completely neglected in the scalar potential,
the VEV $w$ of $\Sigma$ is  then entirely determined by the VEV $v$ of
$\Phi$ and  the quartic couplings $\lambda_\Sigma$  and $\delta$, {\it
viz.}
\begin{equation}
  \label{wtov}
w\ =\ \sqrt{\frac{\delta}{\lambda_\Sigma}}\ v\; .
\end{equation}
This  is an interesting  scenario, since  the ratio  $t_\beta =  v/w =
\sqrt{\lambda_\Sigma/\delta}$   does  not   strongly  depend   on  the
temperature $T$,  as opposed to what  happens to the VEVs  $v$ and $w$
individually.  In fact, as  long as $\lambda_{\Phi,\Sigma}, \delta \ll
1$, the thermally-corrected effective  potential can be expanded, to a
very  good approximation,  in  powers of  $T^2/m^2_\Phi$.   In such  a
high-$T$ expansion, the quartic couplings  of ${\cal L}_V$ turn out to
be $T$-independent~\cite{JKapusta} and hence $t_\beta$ does not depend
on $T$.

We now turn our attention to  the neutrino Yukawa sector of the model,
which is non-standard.  After SSB, it is given in the unitary gauge by
\begin{equation}
  \label{LYuk}
-\, {\cal L}_Y \ =\ \frac{\phi}{v}\ \bar{\nu}_{iL}\, (m_D)_{i
  \alpha}\, \nu_{\alpha R}\
+\ \frac{\sigma\: +\: iJ}{2\,w}\
\bar{\nu}^C_{\alpha R}\, (m_M)_{\alpha\beta}\, 
\nu_{\beta R}\quad +\quad \mbox{H.c.},
\end{equation} 
where summation over repeated  indices is understood. Hereafter we use
Latin indices to label the left-handed neutrinos, e.g.~$\nu_{iL}$, and
Greek  indices  for  the  right-handed  ones,  e.g.~$\nu_{\alpha  R}$.
Observe   that  the   spontaneous  breaking   of   U(1)$_l$  generates
lepton-number-violating    $\Delta   L    =    2$   Majorana    masses
$(m_M)_{\alpha\beta}$  in  addition  to  the  lepton-number-preserving
$\Delta L = 0$ Dirac masses $(m_D)_{i\alpha}$.

The  model under discussion  predicts a  number $(3  + {\rm  n}_R)$ of
Majorana neutrinos  which we collectively  denote by $n_I$, with  $I =
i\,  ,\  \alpha$.   Their   physical  masses  are  obtained  from  the
diagonalisation of the neutrino mass matrix
\begin{equation}
  \label{Mnu}
{\cal M}^\nu \ =\  \left( \begin{array}{cc} 0 & m_D \\
                              m_D^T & m_M \end{array} \right)\; ,
\end{equation}
by means  of the unitary transformation $U^{\nu\,  T}\, {\cal M}^\nu\,
U^\nu =  {\cal \widehat{M}}^\nu$, where ${\cal  \widehat{M}}^\nu$ is a
non-negative diagonal matrix.  The neutrino mass eigenstates $(n_I)_R$
and $(n_I)_L$  are related  to the states  $\nu_{iL}$, $(\nu_{iL})^C$,
$\nu_{\alpha R}$ and $(\nu_{\alpha R})^C$  through
\begin{equation}
  \label{Unu}
\left( \begin{array}{c} \nu_L^C \\ \nu_R \end{array} \right)_I \ =\ 
U^\nu_{IJ}\ (n_J)_R \ ,\qquad
\left( \begin{array}{c} \nu_L \\ \nu^C_R \end{array} \right)_I \ =\ 
U^{\nu\ast}_{IJ}\ (n_J)_L\ .
\end{equation}
Assuming the seesaw hierarchy $(m_D)_{i\alpha}/(m_M)_{\alpha\beta} \ll
1$, the  model predicts  3 light states  that are identified  with the
observed light neutrinos  ($n_i \equiv \nu_i$), and a  number n$_R$ of
heavy Majorana  neutrinos ($n_\alpha \equiv N_\alpha$)  with masses of
order   $(m_M)_{\alpha\beta}    =   \rho_{\alpha\beta}\,   w$,   where
$\rho_{\alpha\beta} = \rho_{\beta\alpha}$  are the Yukawa couplings of
$\Sigma$ to right-handed neutrinos.

To obtain  an accurate light  and heavy neutrino mass  spectrum within
the context of models of electroweak  RL, it is important to go beyond
the  leading seesaw  approximation.  To  this  end, we  need first  to
perform a block diagonalisation and cast ${\cal M}^\nu$ into the form:
\begin{equation}
  \label{block}
{\cal M}^\nu\ \to \ \left( \begin{array}{cc} {\bf m}^\nu & 0 \\
                              0 & {\bf m}^N \end{array} \right)\; .
\end{equation}
This   can   be   achieved   by   introducing   the   unitary   matrix
$V$~\cite{KPSmatrix}:
\begin{equation}
  \label{Vxi}
V\ =\ \left(\! \begin{array}{cc} 
({\bf 1}_3\, +\: \xi^*\xi^T)^{-1/2} & 
\xi^* ({\bf 1}_{{\rm n}_R} +\: \xi^T\xi^*)^{-1/2}\\ 
-\xi^T ({\bf 1}_3\, +\: \xi^*\xi^T)^{-1/2} & 
({\bf 1}_{{\rm n}_R} +\: \xi^T \xi^*)^{-1/2}
\end{array} \!\right)\; ,
\end{equation}
where  $\xi$  is  an   arbitrary  $3\times  {\rm  n}_R$  matrix.   The
expressions  $({\bf  1}_3 +  \xi^*\xi^T)^{-1/2}$  and $({\bf  1}_{{\rm
n}_R} + \xi^T  \xi^*)^{-1/2}$ are defined in terms  of a Taylor series
expansion about  the ${\cal N}\times  {\cal N}$ identity  matrix ${\bf
1}_{\cal N}$. These infinite  series converge provided the norm $||\xi
||$ is much smaller than  1, where $||\xi || \equiv \sqrt{{\rm Tr}(\xi
\xi^\dagger  )}$.  This  condition is  naturally fulfilled  within the
seesaw  framework~\cite{Minkowski}.    Block  diagonalisation  of  the
matrix ${\cal M}^\nu$ given  in~(\ref{Mnu}) implies that the $\{ 12\}$
block element of $V^T {\cal M}^\nu V$ vanishes, or equivalently that
\begin{equation}
  \label{xicondition}
m_D\: -\: \xi\, m_M\: -\: \xi\,m^T_D\,\xi^*\ =\ 0\; .
\end{equation}
Equation~(\ref{xicondition})  determines $\xi$ in  terms of  $m_D$ and
$m_M$.  It can  be solved iteratively, with the  first iteration given
by
\begin{equation}
  \label{xi}
\xi\ =\ m_D\, m_M^{-1}\: -\:
m_D\,m_M^{-1}\,m^T_D\,m^*_D\,m^{*\,-1}_M\, m^{-1}_M\; .
\end{equation}
Note that  the second term on  the RHS of~(\ref{xi})  is suppressed by
the ratio of the light-to-heavy neutrino masses and can thus be safely
neglected  in  numerical estimates.  Upon  block diagonalisation,  the
block mass ``eigen-matrices'' are
\begin{eqnarray}
  \label{blockmN}
{\bf m}^N \!&=&\! \Big({\bf 1}_{{\rm n}_R} +\: \xi^\dagger\xi\Big)^{-1/2}\, 
\Big( m_M\: +\: m^T_D\,\xi^*\: +\: \xi^\dagger m_D \Big)\, 
\Big({\bf 1}_{{\rm n}_R} +\: \xi^T\xi^*\Big)^{-1/2}\; ,\\
  \label{blockmnu}
{\bf m}^\nu \!&=&\! -\, \Big({\bf 1}_3\, +\: \xi \xi^\dagger\Big)^{-1/2}
\Big( m_D\xi^T\: +\: \xi m_D^T\: -\: \xi m_M \xi^T \Big)\,
\Big({\bf 1}_3\, +\: \xi^* \xi^T\Big)^{-1/2}\nonumber\\
\!&=&\! -\: \xi\, {\bf m}^N\,\xi^T\; ,
\end{eqnarray}
where  we  used~(\ref{xicondition}) to  arrive  at  the last  equality
of~(\ref{blockmnu}).  Keeping the leading  order terms in an expansion
of ${\bf m}^N$ in powers of $m_D m^{-1}_M$, we find that
\begin{equation}
  \label{bfMass}
{\bf m}^N\ =\ m_M\: +\: \frac{1}{2}\, \Big(\, 
m_D^\dagger\, m^{-1}_M\, m_D\: +\:
m_D^T\, m^{-1}_M\, m^*_D\, \Big)\; ,\qquad
{\bf m}^\nu\ =\ -\, m_D\, m^{-1}_M\, {\bf m}^N\, m^{-1}_M\, m_D^T\; .
\end{equation}
These  last expressions  are used  to  calculate the  light and  heavy
neutrino   mass   spectra   of   the   RL   scenarios   discussed   in
Section~\ref{sec:EWPT}.

In order to calculate the leptonic asymmetries in the next section, we
need  to know  the Lagrangians  that  govern the  interactions of  the
Majorana  neutrinos $n_I$  and  charged leptons  $l =  e,\,\mu,\,\tau$
with:  ({\it  i})  the  $W^\pm$  and  $Z$  bosons;  ({\it  ii})  their
respective would-be Goldstone bosons  $G^\pm$ and~$G$; ({\it iii}) the
CP-odd Majoron particle  $J$; ({\it iv}) the CP-even  Higgs fields $H$
and  $S$.    In  detail,  these  interaction   Lagrangians  are  given
by~\cite{APMaj}
\begin{eqnarray}
  \label{LagW}
{\cal L}_{W^\mp} \!& = &\! -\, \frac{g_w}{\sqrt{2}}\; W^{-\mu}\  
\bar{l}\, B_{lI}\, \gamma_\mu\,P_L\ n_I \quad + \quad \mbox{H.c.},\\[3mm]
  \label{LagZ}
{\cal L}_Z \!& = &\! -\, \frac{g_w}{4\cos\theta_w}\;  Z^\mu\
\bar{n}_I\, \gamma_\mu\, \Big(\, C_{IJ}\, P_L\: -\: C^*_{IJ}\, P_R\,
\Big)\, n_J\; ,\\[3mm]
  \label{LagGplus}
{\cal L}_{G^\pm} \!& = &\! -\, \frac{g_w}{\sqrt{2}\, M_W}\;  G^-\
\bar{l}\, B_{lI}\, \Big(\, m_l\, P_L\: -\: m_I\, P_R\,
\Big)\, n_I\quad +\quad \mbox{H.c.},\\[3mm]
  \label{LagG}
{\cal L}_G \!& = &\! -\, \frac{i\,g_w}{4 M_W}\; G\
\bar{n}_I\, \Bigg[\, C_{IJ}\, \Big( m_I\, P_L - m_J\, P_R \Big)\: +\:
C^*_{IJ}\, \Big( m_J\, P_L - m_I\, P_R \Big)\, \Bigg]\, n_J\; ,\\[3mm]
  \label{LagJ}
{\cal L}_J \!& = &\! -\, \frac{i\,g_w}{4 M_W}\; t_\beta\, J\
\bar{n}_I\, \Bigg[\, C_{IJ}\, \Big( m_I\, P_L - m_J\, P_R\Big)\: +\:
C^*_{IJ}\, \Big( m_J\, P_L - m_I\, P_R \Big)\nonumber\\
\!&&\! +\: \delta_{IJ}\, m_I \gamma_5\, \Bigg]\; n_J\;,\\[3mm]
  \label{LagH}
{\cal L}_H \!& = &\! -\, \frac{g_w}{4 M_W}\; (c_\theta-s_\theta t_\beta)\;
H\
\bar{n}_I\, \Bigg[\, C_{IJ}\, \Big( m_I\, P_L + m_J\, P_R\Big)\: +\:
C^*_{IJ}\, \Big( m_J\, P_L + m_I\, P_R \Big)\nonumber\\
\!&&\! -\:  
\frac{i\,t_\beta }{t^{-1}_\theta - t_\beta}\  \delta_{IJ}\,
m_I\,\gamma_5\, \Bigg]\; n_J\;,\\[3mm]
  \label{LagS}
{\cal L}_S \!& = &\! -\, \frac{g_w}{4 M_W}\; (s_\theta+c_\theta t_\beta)\;
S\ 
\bar{n}_I\, \Bigg[\, C_{IJ}\, \Big( m_I\, P_L + m_J\, P_R\Big)\: +\:
C^*_{IJ}\, \Big( m_J\, P_L + m_I\, P_R \Big)\nonumber\\
\!&&\! +\:  
\frac{i\,t_\beta }{t_\theta + t_\beta}\  \delta_{IJ}\,
m_I\,\gamma_5\, \Bigg]\; n_J\;,
\end{eqnarray} 
where  $P_{L,R}  =  \frac{1}2\,  (1  \mp \gamma_5  )$,  $g_w$  is  the
 SU(2)$_L$ gauge coupling of the SM and
\begin{equation}
  \label{BC}
B_{l I}\ =\ V^l_{lk}\, U^{\nu\ast}_{kI}\; ,\qquad
C_{IJ}\ =\ U^\nu_{kI}\,U^{\nu\ast}_{kJ}\  .
\end{equation} 
In~(\ref{BC})  $V^l$ is  a 3-by-3  unitary matrix  that occurs  in the
diagonalisation   of    the   charged   lepton    mass   matrix~${\cal
M}^l$. Without  loss of generality,  we assume throughout  the present
study that ${\cal  M}^l$ is positive and diagonal,  which implies that
$V^l = {\bf  1}_3$.  Finally, we comment on the  limit of $t_\beta \to
0$.  It is  easy to see from~(\ref{t2theta}) that  this limit leads to
$t_\theta \to 0$ and the fields $S$ and $J$ decouple from matter; only
the Higgs field  $H$ couples to Majorana neutrinos and  to the rest of
the SM fermions~(cf.~\cite{APZPC}).

\setcounter{equation}{0}
\section{Leptonic Asymmetries}\label{sec:asym}

In this section we calculate  the leptonic asymmetries produced by the
decays  of the heavy  Majorana neutrinos  during a  second-order EWPT.
The novel aspect  of such a calculation is that,  in stark contrast to
the  conventional leptogenesis  scenario, the  $W^\pm$ and  $Z$ bosons
also contribute  to the decays  and leptonic asymmetries of  the heavy
Majorana neutrinos.  This fact raises  new issues related to the gauge
invariance  of  off-shell Green  functions  which  are here  addressed
within the so-called Pinch Technique (PT) framework~\cite{PT}.

Since  sphalerons act  on the  left-handed SM  fermions  converting an
excess in leptons  into that of baryons, we only  need to consider the
decays of the heavy Majorana neutrinos $N_\alpha$ into the left-handed
charged leptons $l^-_L$ and light neutrinos $\nu_{lL}$.  In detail, we
have  to calculate  the  partial  decay width  of  the heavy  Majorana
neutrino $N_\alpha$ into a particular lepton flavour $l$,
\begin{equation}
  \label{GammaN}
\Gamma^l_{N_\alpha}\ =\ \Gamma (N_\alpha \to l^-_L\ W^+,\, G^+ )\ +\ 
\Gamma (N_\alpha \to \nu_{lL}\ Z,\, G,\, J,\, H,\, S )\ . 
\end{equation}
To compute  $\Gamma^l_{N_\alpha}$, it proves more  convenient to first
calculate   the   absorptive   part  $\Sigma^{\rm   abs}_{\alpha\beta}
(\not\!p)$  of  the  heavy  Majorana-neutrino  self-energy  transition
$N_\beta \to N_\alpha$ in the Feynman--'t Hooft gauge $\xi = 1$, where
$p^\mu$  is   the  4-momentum  carried   by  $N_{\alpha,\beta}$.   The
Feynman--'t  Hooft gauge  is not  a simple  choice of  gauge,  but the
result   obtained   in   the   gauge-independent  framework   of   the
PT~\cite{PT}, within  which issues  of analyticity, unitarity  and CPT
invariance   can   self-consistently   be   addressed~\cite{PP,APNPB}.

%******************************************************************
%%% Figure 1 on self-energy transitions
%******************************************************************
\begin{figure}[t]
\begin{center}
\begin{picture}(350,100)(0,0)
\SetWidth{0.8}

\ArrowLine(0,50)(30,50)\ArrowLine(30,50)(90,50)\ArrowLine(90,50)(120,50)
\PhotonArc(60,50)(30,0,180){3}{6.5}
\Text(5,45)[t]{$N_\beta$}\Text(60,45)[t]{$l^-_L\,,\ \nu_{lL}$}
\Text(120,45)[t]{$N_\alpha$}
\Text(60,87)[b]{$W^+,\ Z$}

\Text(60,10)[]{\bf (a)}

\ArrowLine(200,50)(230,50)\ArrowLine(230,50)(290,50)
\ArrowLine(290,50)(320,50)\DashArrowArcn(260,50)(30,180,0){4}
\Text(205,45)[t]{$N_\beta$}\Text(260,45)[t]{$l^-_L\,,\ \nu_{lL}$}
\Text(320,45)[t]{$N_\alpha$}
\Text(260,87)[b]{$G^+,\ G,\ J,\ H,\ S$}

\Text(260,10)[]{\bf (b)}

\end{picture}
\end{center}
\caption{\it Feynman graphs that determine the 1-loop absorptive part
  $A_{\alpha\beta} (s)$ of the heavy Majorana-neutrino
  self-energy $\Sigma_{\alpha\beta} (\not\! p)$.}\label{fig:abs}
\end{figure}
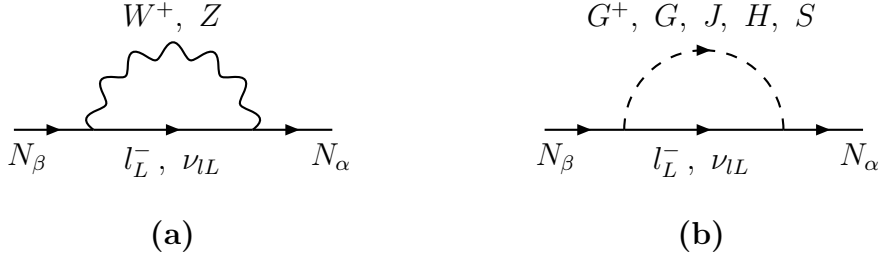

Neglecting  the   small  charged-lepton  and   light-neutrino  masses,
$\Sigma^{\rm   abs}_{\alpha\beta}  (\not\!p)$   acquires   the  simple
spinorial structure:
\begin{equation}
  \label{Selfabs}
\Sigma^{\rm abs}_{\alpha\beta}  (\not\!  p)\ =\ A_{\alpha\beta} (s)
\not\!  p\, P_L\ +\ A^*_{\alpha\beta} (s)
\not\!  p\, P_R\; , 
\end{equation}
where $s  = p^2$ is  the squared Lorentz-invariant mass  associated to
the  self-energy transition $N_\beta  \to N_\alpha$.   Considering the
Feynman  graphs  shown   in  Fig.~\ref{fig:abs}  and  the  interaction
Lagrangians~(\ref{LagW})--(\ref{LagS}),   the   absorptive  transition
amplitudes $A_{\alpha\beta} (s)$ are calculated to be
\begin{eqnarray}
  \label{Abs}
A_{\alpha\beta} (s) \!& = &\! \frac{\alpha_w}{32}\,
\sum\limits_{l = e,\mu ,\tau} 
\Bigg\{ B^*_{l\alpha}B_{l\beta}\,\Bigg[\, 
4\,\Bigg( 1 - \frac{M^2_W}{s} \Bigg)^2 \theta (s - M^2_W )\ +\
\frac{2\,M^2_Z}{M^2_W}\; \Bigg( 1 - \frac{M^2_Z}{s} \Bigg)^2
\theta (s - M^2_Z )\,\Bigg]\nonumber\\
\!&&\! \hspace{-1.2cm}
+\; 
\frac{m_{N_\alpha}\, m_{N_\beta}}{M^2_W}\, 
B_{l\alpha}B^*_{l\beta}\,
\Bigg[\, 2\, \Bigg( 1 -
\frac{M^2_W}{s} \Bigg)^2 \theta (s - M^2_W )\ +\
\Bigg( 1 - \frac{M^2_Z}{s} \Bigg)^2 \theta (s - M^2_Z )\ +\ t^2_\beta\,
\theta (s)\nonumber\\
\!&&\! \hspace{-1.2cm}
+\, (c_\theta - s_\theta t_\beta )^2 \Bigg( 1 - \frac{M^2_H}{s}
\Bigg)^2 \theta (s - M^2_H )\ +\ 
 (s_\theta + c_\theta t_\beta )^2 \Bigg( 1 - \frac{M^2_S}{s}
\Bigg)^2 \theta (s - M^2_S )\; \Bigg]\, \Bigg\}\; ,
\end{eqnarray}
where  $\alpha_w  =  g^2_w/(4\pi)$  is  the  SU(2)$_L$  fine-structure
constant and $\theta (x)$ is the usual step function: $\theta (x) = 1$
for $x >0$, whilst $\theta (x) = 0$ if $x \leq 0$.  In the calculation
of  $A_{\alpha\beta} (s)$,  we  used  the fact  that  $B_{l \alpha}  =
C_{\nu_l  \alpha}  +  {\cal   O}(C^2_{\nu_l  \alpha})$,  which  is  an
excellent  approximation in  the physical  charged-lepton  mass basis.

We  should  bear  in  mind   that  all  masses  involved  on  the  RHS
of~(\ref{Abs})   depend   on   the   temperature  $T$,   through   the
$T$-dependent  VEVs $v(T)$  and $w(T)$  related to  the  Higgs doublet
$\Phi$ and the  complex singlet $\Sigma$, respectively [cf.~(\ref{vT})
and~(\ref{wT})].   In  the symmetric  phase  of  the theory,  i.e.~for
temperatures above the electroweak phase transition, these VEVs vanish
and the absorptive transition amplitude becomes
\begin{equation}
  \label{AbsSymmetric}
A_{\alpha\beta} (s) \ =\ 
\frac{\alpha_w}{8}\
 \frac{(m^T_D\, m^*_D)_{\alpha\,\beta}}{M^2_W}\ 
\Bigg(\, 1\: +\:
\frac{t^2_\beta}{2}\;\Bigg)\; .
\end{equation}
Note that this  last formula is only valid in the  weak basis in which
the Majorana mass matrix $m_M$ is diagonal.

To account for unstable-particle-mixing effects between heavy Majorana
neutrinos, we follow~\cite{APRD,PU}  and define the resummed effective
couplings   $\overline{B}_{l\alpha}$  and   their   CP-conjugate  ones
$\overline{B}^c_{l\alpha}$ related  to the vertices  $W^-l_L N_\alpha$
and $W^+ (l_L)^C N_\alpha$, respectively. For a symmetric model with 3
left-handed  and  3 right-handed  neutrinos,  the effective  couplings
$\overline{B}_{l\alpha}$ exhibit  the same analytic  dependence on the
absorptive  transition amplitudes $A_{\alpha\beta}$  as the  one found
in~\cite{PU}:\footnote{Here   we  eliminate   a   typo  that   occurred
in~\cite{PU},  where  $R_{\alpha\gamma}$   in  the  numerator  of  the
fraction needs be multiplied with $-i$.}
\begin{eqnarray}
  \label{hres3g}
\overline{B}_{l\alpha} \!&=&\!  B_{l\alpha}\: -\: i\,
\sum\limits_{\beta,\gamma = 1}^3 
|\varepsilon_{\alpha\beta\gamma}|\; B_{l\beta}\\
&&\hspace{-1.35cm}\times\,\frac{m_\alpha ( m_\alpha A_{\alpha\beta} +
  m_\beta A_{\beta\alpha}) - i R_{\alpha \gamma} \Big[ m_\alpha
    A_{\gamma\beta} ( m_\alpha A_{\alpha\gamma} + m_\gamma A_{\gamma\alpha}
    ) + m_\beta A_{\beta\gamma} ( m_\alpha A_{\gamma\alpha} + m_\gamma
    A_{\alpha \gamma} ) \Big]} { m^2_\alpha\, -\, m^2_\beta\, +\,
  2i\,m^2_\alpha A_{\beta\beta} + 2i\,{\rm Im}R_{\alpha\gamma}\, \Big(
  m^2_\alpha |A_{\beta\gamma}|^2 + m_\beta m_\gamma {\rm
    Re}A^2_{\beta\gamma}\Big) }\ ,\nonumber
\end{eqnarray}
where  all transition amplitudes  $A_{\alpha\beta}$, $A_{\beta\gamma}$
etc are evaluated at $s = m^2_{N_\alpha} \equiv m^2_\alpha$ and
\begin{equation}
R_{\alpha \beta}\ =\ \frac{m^2_\alpha}{m^2_\alpha - m^2_\beta + 
2i\, m^2_\alpha A_{\beta\beta} (m^2_\alpha)}\ .
\end{equation}
Moreover,  $|\varepsilon_{\alpha\beta\gamma}|$ is  the modulus  of the
usual Levi--Civita anti-symmetric tensor.  The respective CP-conjugate
effective   couplings   $\overline{B}^c_{li}$   are  easily   obtained
from~(\ref{hres3g})  by replacing  the ordinary  $W^-$-boson couplings
$B_{l\alpha}$ and  $A_{\alpha\beta} (s)$ by  their complex conjugates.
In the decoupling  limit of $m_{N_3} \gg m_{N_{1,2}}$,  we recover the
analytic   results   known   for   a   model   with   2   right-handed
neutrinos~\cite{APRD,PU},     where     the    effective     couplings
$\overline{B}_{l1,2}$ are given by
\begin{eqnarray}
  \label{B2gen1}
\overline{B}_{l1} \!& = &\! B_{l1}\ -\ i\, B_{l2}\, \frac{m_{N_1}\,\Big(\, 
m_{N_1}\, A_{12} (m^2_{N_1})\: +\: m_{N_2}\, A_{21} (m^2_{N_1})\, \Big)}
{ m^2_{N_1}\: -\: m^2_{N_2}\ + \ 2i m^2_{N_1}\, A_{22} (m^2_{N_1})}\ ,\\[3mm]
  \label{B2gen2}
\overline{B}_{l2} \!& = &\! B_{l2}\ -\ i\, B_{l1}\, \frac{m_{N_2}\,
\Big(\, m_{N_2}\, A_{21} (m^2_{N_2})\: +\: 
m_{N_1}\, A_{12} (m^2_{N_2})\, \Big)}
{ m^2_{N_2}\: -\: m^2_{N_1}\ + \ 2i m^2_{N_2}\, A_{11} (m^2_{N_2})}\ .
\end{eqnarray}
In all our results, we  neglect the 1-loop corrections to the vertices
$W^\pm l_L N_\alpha$,  $Z\nu_{lL}N_\alpha$ etc, whose absorptive parts
are numerically insignificant in leptogenesis, but essential otherwise
to   ensure   gauge   invariance   and   unitarity   within   the   PT
framework~\cite{APNPB}.

In terms of  the resummed effective couplings $\overline{B}_{l\alpha}$
and   $\overline{B}^c_{l\alpha}$   and   the   absorptive   transition
amplitudes   $A_{\alpha\beta}   (s)$,   the   partial   decay   widths
$\Gamma^l_{N_\alpha       }$       and       their       CP-conjugates
$\overline{\Gamma}^l_{N_\alpha }$ are now given by
\begin{equation}
  \label{Widths}
\Gamma^l_{N_\alpha }\ =\  m_{N_\alpha}\, A_{\alpha\alpha}
      ( m^2_{N_\alpha};\, \overline{B}_{l\alpha})\; ,\qquad
\overline{\Gamma}^{\; l}_{N_\alpha }\ =\  m_{N_\alpha}\, A_{\alpha\alpha}
      (m^2_{N_\alpha};\, \overline{B}^c_{l\alpha})\; ,
\end{equation}
where  the  dependence  of  the absorptive  transition  amplitudes  on
$\overline{B}_{l\alpha}$ and~$\overline{B}^c_{l\alpha}$ has explicitly
been indicated.   Note that no  summation over the  individual charged
leptons and  light neutrinos running  in the loop should  be performed
when  calculating  $\Gamma^l_{N_\alpha  }$ and  $\overline{\Gamma}^{\;
l}_{N_\alpha  }$  using~(\ref{Abs})  and  (\ref{Widths}).   Then,  the
leptonic asymmetries  for each  individual lepton flavour  are readily
found to be
\begin{equation}
  \label{deltaN}
\delta^l_{N_\alpha}\ =\ \frac{ \Delta \Gamma^l_{N_\alpha} }{ 
\Gamma_{N_\alpha} } \ =\ 
\frac{ |\overline{B}_{l\alpha}|^2\: -\: |\overline{B}^c_{l\alpha}|^2}
{\sum\limits_{l = e,\mu ,\tau} 
\Big(\, |\overline{B}_{l\alpha}|^2\: +\:
  |\overline{B}^c_{l\alpha}|^2\,\Big) }\ ,
\end{equation}
with 
\begin{equation}
\Gamma_{N_\alpha} \ =\ \sum\limits_{l = e,\mu ,\tau}\,
\Big(\, \Gamma^l_{N_\alpha}\: +\: 
\overline{\Gamma}^{\; l}_{N_\alpha}\,\Big)\;, \qquad
\Delta\Gamma^l_{N_\alpha} \ =\ \Gamma^l_{N_\alpha}\: -\: 
\overline{\Gamma}^{\; l}_{N_\alpha}\; .
\end{equation}
Notice  that  both  $\Gamma_{N_\alpha}$  and  $\delta^l_{N_i}$  do  in
general  depend  on the  temperature  $T$,  through the  $T$-dependent
masses,  during  a  second-order  electroweak phase  transition.  More
details on this issue will be presented in the next section.

\setcounter{equation}{0}
\section{Electroweak Resonant Leptogenesis}\label{sec:EWPT}

In this section we present the relevant Boltzmann equations (BEs) that
will enable  us to evaluate the  lepton-to-photon and baryon-to-photon
ratios, $\eta_{L_l}$ and $\eta_B$, during a second-order EWPT.  In our
numerical  estimates, we  only  include the  dominant collision  terms
related to the $1 \leftrightarrow  2$ decays and inverse decays of the
heavy  Majorana  neutrinos   $N_\alpha$.   We  also  neglect  chemical
potential  contributions  from the  right-handed  charged leptons  and
quarks~\cite{reviews}.   A  complete  account  of  the  aforementioned
subdominant effects may be given elsewhere.

To start  with, we  first write  down the BEs  that govern  the photon
normalised number densities  $\eta_{N_\alpha}$ and $\eta_{\Delta L_l}$
for  the  heavy  Majorana  neutrinos $N_\alpha$  and  the  left-handed
leptons $l_L$, $\nu_{lL}$, respectively:
\begin{eqnarray}
  \label{BEN}
\frac{d\eta_{N_\alpha}}{dz} \!& =&\! \frac{z\, D_{N_\alpha}}{H(T_c)}\ 
\Bigg(\, 1\: -\: \frac{\eta_{N_\alpha}}{\eta^{\rm eq}_{N_\alpha}}\,
  \Bigg)\; ,\\
  \label{BEDL}
\frac{d\eta_{\Delta L_l}}{dz} \!& =&\! \frac{z\, D_{N_\alpha}}{H(T_c)}\ 
\Bigg[\, \Bigg(\, \frac{\eta_{N_\alpha}}{\eta^{\rm eq}_{N_\alpha}}\: -\: 1\,
  \Bigg)\, \delta^l_{N_\alpha}\ 
  -\ \frac{2}{3}\; B^l_{N_\alpha}\, \eta_{\Delta L_l}\,
  \Bigg] \; .
\end{eqnarray}
Although  our conventions  and notations  follow  those of~\cite{PU2},
there are several key differences  pertinent to our EWPT scenario that
need to be stressed here.  Specifically, we express the $T$-dependence
of the BEs~(\ref{BEN}) and~(\ref{BEDL})  in terms of the dimensionless
parameter $z$:
\begin{equation}
  \label{zparam}
z\ =\ \frac{T_c}{T}\ ,
\end{equation}
where $T_c$ is  the critical temperature of the  EWPT to be determined
below   [cf.~(\ref{Tc})].   The   parameter   $H(T_c)\approx  17\times
T^2_c/M_{\rm P}$ is the Hubble constant at $T=T_c$, where $M_{\rm P} =
1.2\times   10^{19}$~GeV   is   the   Planck  mass.    The   parameter
$B^l_{N_\alpha}$ denotes  the branching fraction of the  decays of the
heavy Majorana  neutrino $N_\alpha$  into a particular  lepton flavour
$l$,     i.e.~$B^l_{N_\alpha}     =     (\Gamma^l_{N_\alpha}\,     +\,
\overline{\Gamma}^{\;   l}_{N_\alpha})/\Gamma_{N_\alpha}$.   Moreover,
$\eta_{N_\alpha}^{\rm eq}$  is the  equilibrium number density  of the
heavy neutrino $N_\alpha$, normalised to the number density of photons
$n_\gamma = 2T^3/\pi^2$:
\begin{equation}
  \label{etaNeq}
\eta_{N_\alpha}^{\rm eq} \ =\ \frac{m^2_{N_\alpha}(T)}{2T^2}\; 
K_2\Bigg(\frac{m_{N_\alpha} (T)}{T}\Bigg)\; ,
\end{equation}
where $K_n (x)$ is the $n$th-order modified Bessel function~\cite{AS}.
Finally, $D_{N_\alpha}$ is the $T$-dependent collision term related to
the decay and inverse decay of the heavy Majorana neutrino~$N_\alpha$:
\begin{equation}
  \label{DNalpha}
D_{N_\alpha}\      =\     \frac{\Gamma_{N_\alpha}     (T)}{n_\gamma}\;
g_{N_\alpha}\,                   \int\,                  \frac{d^3{\bf
p}_{N_\alpha}}{(2\pi)^3}\,\frac{m_{N_\alpha}(T)}{E_{N_\alpha}(T)}\,
e^{-E_{N_\alpha}(T)/T}    \   =\   \frac{m^2_{N_\alpha}(T)}{2T^2}\,
\Gamma_{N_\alpha}     (T)\, K_1\Bigg(\frac{m_{N_\alpha} (T)}{T}\Bigg)\; ,
\end{equation}
where    $E_{N_\alpha}(T)    =    [\,|{\bf   p}_{N_\alpha}|^2\:    +\:
m^2_{N_\alpha}(T)\,]^{1/2}$ and  $g_{N_\alpha } = 2$ is  the number of
helicities of $N_\alpha$.

Our  next step  is  to  include the  effect  of the  $(B+L)$-violating
sphalerons~\cite{KRS} on  the lepton-number densities  produced by the
decays of $N_\alpha$  during the EWPT.  In particular,  our interest is
to implement the temperature dependence of the rate of $B+L$ violation
just  below  the critical  temperature  $T_c$,  where  $T_c$ is  given
by~\cite{MEC}
\begin{equation}
  \label{Tc}
T_c\ =\ v\, \left(\,\frac{1}{2}\: +\:
\frac{3\,g^2_w}{8\,\lambda_\Phi}\: +\:
\frac{g^{\prime\,2}}{8\,\lambda_\Phi}\: +\:
\frac{h^2_{t}}{2\,\lambda_\Phi}\, \right)^{-1/2} .
\end{equation}
In the above,  $g^\prime$ is the U(1)$_Y$ gauge  coupling and $h_t$ is
the top-quark Yukawa coupling.   We should notice that $\Phi$-$\Sigma$
mixing  effects  have been  omitted  in~(\ref{Tc}),  which  is a  good
approximation for  scenarios with $\delta /\lambda_\Phi \ll  1$ as the
ones to be considered here.

A  reliable estimate~\cite{AM,CLMW} of  the rate  of $(B+L)$-violating
sphaleron transitions can be  obtained for temperatures satisfying the
double inequality
\begin{equation}
  \label{BLcond}
M_W(T)\ \ll\ T\ \ll\ \frac{M_W(T)}{\alpha_w}\;,
\end{equation}
where  $\alpha_w  =  g^2_w/4\pi$   is  the  SU(2)$_L$  fine  structure
constant, $M_W(T)  = g_w\,v(T)/2$ is the  $T$-dependent $W$-boson mass
and
\begin{equation}
  \label{vT}
v(T)\ =\ v\, \left(\,1\: -\: \frac{T^2}{T_c^2}\,\right)^{1/2}
\end{equation}
is the  $T$-dependent VEV of the  Higgs field. In detail,  the rate of
$B+L$ violation per unit volume is~\cite{AM}
\begin{equation}
  \label{BLrate}
\gamma_{\Delta (B+L)}\ =\
\frac{\omega_-}{2\,\pi}\;{\cal N}_{\rm tr}\, ({\cal N}V)_{\rm rot}\,
\left(\frac{\alpha_w\,T}{4\,\pi} \right)^3 \alpha_3^{-6}\,e^{-E_{\rm
sp} / T}\, \kappa\; .
\end{equation}
Given the  double inequality  (\ref{BLcond}), this last  expression is
valid for  temperatures $T \stackrel{<}{{}_\sim}  T_c$.  Following the
notation of~\cite{AM}, the  parameters $\omega_-$, ${\cal N}_{\rm tr}$
and ${\cal N}_{\rm rot}$ that occur in~(\ref{BLrate}) are functions of
$\lambda_\Phi  /  g^2_w$,  $V_{\rm  rot}  = 8\pi^2$  and  $\alpha_3  =
\alpha_w\,T/[2\,M_W(T)]$.    The   quantity   $E_{\rm  sp}$   is   the
$T$-dependent energy of the sphaleron and is determined by
\begin{equation}
E_{\rm sp}\ =\ A\,\frac{2\,M_W(T)}{\alpha_w}\ ,
\end{equation}
where $A$ is a function of  $\lambda_\Phi / g^2_w$ and is ${\cal O}( 1
)$, for  values of phenomenological  interest.  The dependence  of the
parameter  $\kappa$   on  $\lambda_\Phi/g^2_w$  has   been  calculated
in~\cite{AM,CLMW}, and the results  of those studies are summarised in
Table~\ref{BLparams}, for  $\lambda_\Phi /g^2_w =  0.556$.  This value
corresponds to a SM Higgs-boson mass $M_H$ of 120~GeV in the vanishing
limit of a $\Phi$-$\Sigma$ mixing.

\begin{table}[t]
\begin{center}
\begin{tabular}{|c||c|c|c|c|c|}
\hline &&&&&\\[-11pt] 
$\lambda_\Phi / g^2_w$ & $\omega_-$ & ${\cal N}_{\rm
rot}$ & ${\cal N}_{\rm tr}$ & $\kappa$ & $A$\\[2pt] \hline \hline
0.556 & 1.612$\times M_W$ & 11.2 & 7.6 & 0.135 -- 1.65 & 1\\ \hline
\end{tabular}
\end{center}
\caption{\it Values of the  parameters occurring in (\ref{BLrate}) for
$\lambda_\Phi/g^2_w =  0.556$, which  corresponds to a  SM Higgs-boson
mass of 120~GeV when $\delta = 0$.}\label{BLparams}
\end{table}

Since the SM Higgs-boson  mass is $M_H \stackrel{>}{{}_\sim} 115$~GeV,
it  can be  shown~\cite{EWSM} that  the EWPT  in the  SM is  not first
order,  but   continuous  from  $v(T_c)=0$  to   $v$,  without  bubble
nucleation  and  the formation  of  large  spatial inhomogeneities  in
particle  densities.   Therefore,   we  use  the  formalism  developed
in~\cite{PU2},  where  the  $(B+L)$-violating  sphaleron  dynamics  is
described  in  terms  of  spatially independent  $B$-  and  $L$-number
densities  $\eta_{B}$ and~$\eta_{L_j}$.  More  explicitly, the  BEs of
interest to us are~\cite{PU2}:
\begin{eqnarray}
  \label{Bsph}
\frac{d \eta_{B}}{dz} \!& = &\! -\, \frac{z\,\Gamma_{\Delta
    (B+L)}}{H(T_c)}\; \bigg[\, \eta_{B}\: +\: 
\frac{28}{51}\, \eta_L\: 
+\: \frac{v^2(T)}{T^2}\,
\bigg(\, \frac{75}{187}\, \eta_{B}\: +\:\frac{16}{187}\, 
\eta_L\,\bigg)\,\bigg]\; ,\qquad\\[12pt]
  \label{Lsph}
\frac{d \eta_{L_i}}{dz} \!& = &  \!
\frac{d\eta_{\Delta L_i}}{dz}
\: +\: \frac{1}{3}\,
\frac{d \eta_{B}}{dz}\ ,
\end{eqnarray}
where  $\eta_L = \sum_{l\,=e,\mu,\tau} \eta_{L_l}$  is the  total
lepton asymmetry and
\begin{equation}
\Gamma_{\Delta (B+L)}\ =\ \frac{1683}{
132\, T^3\: +\: 51\, T\,v^2(T) }\  \gamma_{\Delta (B+L)} \; .
\end{equation}
We observe that in the limit $\Gamma_{\Delta (B+L)}/H(T_c) \to \infty$
and  for $T  >  T_c$,  the conversion  of  the lepton-to-photon  ratio
$\eta_L$ to the baryon-to-photon ratio~$\eta_B$ is given by the known
relation~\cite{KS,LS}:
\begin{equation}
\eta_B\ =\ -\,\frac{28}{51}\: \eta_L\; .
\end{equation}
Likewise, when  $1 \stackrel{<}{{}_\sim} z  \stackrel{<}{{}_\sim} 1.7$
and $\kappa  =1$, it is  $\Gamma_{\Delta (B+L)}/H(T_c) \gg 1$  and the
baryon-to-photon  ratio   $\eta_B$  is  then  related   to  the  total
lepton-to-photon ratio $\eta_L$ by
\begin{equation}
  \label{etaBL}
\eta_B \ = \ -\ \Bigg(\,\frac{28}{51}\: +\: \frac{16}{187}\:
\frac{v^2(T)}{T^2}\, \Bigg)\, \Bigg( 1\: +\: \frac{75}{187}\:
\frac{v^2(T)}{T^2} \Bigg)^{-1}\, \eta_L\; .
\end{equation}
For $z\stackrel{>}{{}_\sim} 1.7$, sphaleron effects get sharply out of
equilibrium and $\eta_B$ freezes out. To account for the $T$-dependent
$(B+L)$-violating sphaleron  effects, our numerical  estimates will be
based   on  the   BEs~(\ref{BEN}),   (\ref{BEDL}),  (\ref{Bsph})   and
(\ref{Lsph}).
 
In the singlet  Majoron model, the restoration of  the global symmetry
U(1)$_l$  will occur  for  temperatures above  a critical  temperature
$T^l_c$ that could in general differ  from $T_c$ of the SM gauge group
given in~(\ref{Tc}).  For example, in the absence of a doublet-singlet
mixing,  the  critical temperature  related  to  the  SSB of  U(1)$_l$
is~\cite{JKapusta}
\begin{equation}
  \label{Tlc}
T^l_c \ =\ -\ \frac{6\, m^2_\Sigma}{\lambda_\Sigma}\ .
\end{equation}
Consequently, the $T$-dependence of $w(T)$ for $T< T^l_c$ will be
analogous to $v(T)$ in~(\ref{vT}), i.e.
\begin{equation}
  \label{wT}
w(T)\ =\ w\, \left(\,1\: -\:
\frac{T^2}{(T_c^l)^2}\,\right)^{1/2}\ .
\end{equation}
However,  if $m^2_\Sigma$  vanishes, the  singlet VEV  $w(T)$  and the
doublet VEV  $v(T)$ will  be related by  an expression  very analogous
to~(\ref{wtov}), namely
\begin{equation}
  \label{wTtovT}
w(T)\ \approx \ t^{-1}_\beta\, v(T)\; .
\end{equation}
As was mentioned after  (\ref{wtov}), the above relation becomes exact
in  a   high-$T$  expansion  of  the   thermally  corrected  effective
potential. Such an expansion is a very good approximation to the level
of     a     few    \%     for     perturbatively    small     quartic
couplings~\cite{JKapusta}. As  a consequence,  the SM gauge  group and
the global lepton symmetry U(1)$_l$ will both break down spontaneously
via the same second-order  electroweak phase transition, with $T^l_c =
T_c$.  Even  though the focus  of the paper  will be on this  class of
scenarios,  we will comment  on possible  differences for  models with
$T^l_c \neq T_c$.

If  $T^l_c  = T_c$,  the  heavy  neutrino  masses $m_{N_\alpha}$,  the
gauge-boson masses $M_{W,Z}$ and  the Higgs masses $M_{H,S}$ all scale
with  the  same  $T$-dependent  factor,  $(1-T^2/T^2_c)^{\,1/2}$,  for
temperatures $T<T_c$ of our interest.  Hence, the $T$-dependence drops
out exactly in the expression~(\ref{Abs}) of the absorptive transition
amplitudes  $A_{\alpha\beta} (m^2_{N_\gamma})$,  and  likewise in  the
leptonic asymmetries $\delta^l_{N_\alpha}$ and the branching fractions
$B^l_{N_\alpha}$.  However,  as can be  seen from~(\ref{DNalpha}), the
collision  terms $D_{N_\alpha}$  exhibit a  non-trivial $T$-dependence
that needs be carefully implemented in the BEs.

For our numerical  estimates of the BAU, we  consider the 3-generation
flavour scenario  of   the  RL  model  discussed  in~\cite{APtau,PU2}.
Specifically, the Majorana sector is assumed to be approximately SO(3)
symmetric,
\begin{equation}
  \label{mM}
m_M\ =\ m_N\, {\bf 1}_3\: +\: \Delta M_S\; ,
\end{equation}
where $\Delta  M_S$ are small  SO(3)-breaking terms that are  of order
$m^\dagger_D\,    m_D/m_N$   as    these   are    naturally   expected
from~(\ref{bfMass}). Plugging~(\ref{mM})  into~(\ref{bfMass}), we find
that, to leading order in $\Delta M_S$, the heavy neutrino mass matrix
${\bf m}^N$ deviates from $ m_N\, {\bf 1}_3$ by an amount
\begin{equation}
  \label{dmN}
\delta {\bf m}^N \ =\ \Delta M_S\: +\: \frac{1}{2 m_N}\;
\Big( m^\dagger_D\, m_D\: +\: m^T_D\, m^*_D \Big)\; .
\end{equation}
It is interesting to  observe that possible renormalisation-group (RG)
running effects  from a high-energy scale $M_X$,  e.g.~GUT scale, down
to  $m_N$  will  induce   a  negative  contribution  to  $\delta  {\bf
m}^N$~\cite{rad}, i.e.
\begin{equation}
(\delta {\bf m}^N)_{\rm RG} \ =\ -\frac{\alpha_w}{8\pi}\;
  \frac{m_N}{M^2_W}\ \Big(\, m^\dagger_D m_D\: +\: m^T_D m_D^*\, \Big)\;
\ln\Bigg( \frac{M_X}{m_N}\Bigg)\; .
\end{equation}
For  $M_X = M_{\rm  GUT} \sim  10^{16}$ and  $m_N =  80$--150~GeV, the
RG-induced  terms are typically  smaller by  a factor  $\sim 0.1$--0.4
with  respect  to the  tree-level  contribution given  in~(\ref{dmN}).
Thus, the  inclusion of  the RG  effects are not  going to  affect the
results of our analysis in a substantial manner.

As was  mentioned already, the SO(3)  symmetry is broken  by the Dirac
mass terms $(m_D)_{i\alpha}$, which in our case possess an approximate
U(1)-symmetric flavour pattern~\cite{APtau}:
\begin{equation}
  \label{mD}
m_D\ =\ \frac{v}{\sqrt{2}}\, 
\left( \begin{array}{ccc}
0 & a\,e^{-i\pi/4} & a\,e^{i\pi/4} \\
0 & b\,e^{-i\pi/4} & b\,e^{i\pi/4} \\
0 & c\,e^{-i\pi/4} & c\,e^{i\pi/4} \\
\end{array}\right)\ +\ \delta m_D\; ,
\end{equation}
where the 3-by-3  matrix $\delta m_D$,
\begin{equation}
  \label{dmD}
\delta m_D\ =\ \frac{v}{\sqrt{2}}\, \left(
\begin{array}{ccc}
\varepsilon_e & 0 & 0\\
\varepsilon_\mu & 0 & 0\\
\varepsilon_\tau & 0 & 0\end{array} \right)\; ,
\end{equation}
violates the  U(1) symmetry  by small terms  of order of  the electron
mass $m_e$.  Instead, the  U(1)-symmetric Yukawa couplings $a$ and $b$
can  be as  large  as the  $\tau$-lepton  Yukawa coupling  $m_\tau/v$,
i.e.~of  order  $10^{-2}$--$10^{-3}$.    For  successful  RL,  it  was
found~\cite{APtau,PU2} that the parameter $c$ needs to be taken of the
order of  the electron  Yukawa coupling $m_e/v$.   It is  important to
stress  here that the  approximate flavour  symmetries SO(3)  and U(1)
ensure  the stability of  the light-  and heavy-neutrino  sector under
loop corrections~\cite{APZPC,APtau,Kersten/Smirnov}.

\begin{table}[t]
\begin{center}
{\small 
\begin{tabular}{|c||c|c|c|c|c|c|}
\hline 
  & & & & & & \\[-3mm]
{\bf Higgs} & $\lambda_\Phi$ & $\lambda_\Sigma$ & $\delta$ & 
$\tan\beta$ & $M_H$~[GeV] & $M_S$~[GeV] \\ 
{\bf Sector}  &  & & & & & \\
\hline  
& & & & & & \\[-4mm]
& 0.238 & $\frac{1}{30}\,\lambda_\Phi$ & $\frac{1}{15}\,\lambda_\Phi$
  & $1/\sqrt{2}$ & 121 & 29 \\[-4mm]
  & & & & & & \\
\hline\hline
  & & & & & & \\[-3mm]
{\bf Neutrino} & 
$\frac{\displaystyle (\delta {\bf m}^N)_{11}}{\displaystyle m_N}$ &
$\frac{\displaystyle (\delta {\bf m}^N)_{12}}{\displaystyle m_N}$ &  
$\frac{\displaystyle (\delta {\bf m}^N)_{13}}{\displaystyle m_N}$ &
$\frac{\displaystyle (\delta {\bf m}^N)_{22}}{\displaystyle m_N}$ &
$\frac{\displaystyle (\delta {\bf m}^N)_{23}}{\displaystyle m_N}$ &
$\frac{\displaystyle (\delta {\bf m}^N)_{33}}{\displaystyle m_N}$ \\
{\bf Sector} & & & & & & \\
\hline
  & & & & & & \\[-4mm]
  & $10^{-5}$ & $-10^{-9}$ & $-4\times 10^{-10}$ & $4\times 10^{-9}$ & 
  $(6.8 - 0.6 i)$ & $5.2\times 10^{-9}$\\
  & & & & & ~~~~~~~$\times 10^{-9}$ & \\
\hline
  & $a$ & $b$ & $c$ & 
            $\varepsilon_e$ & $\varepsilon_\mu$ & $\varepsilon_\tau$\\
\hline  
  & & & & & & \\[-4mm]
  & $\frac{\displaystyle 3}{\displaystyle 500}$ & 
    $\frac{\displaystyle 57}{\displaystyle 25000}$ & $2\times 10^{-7}$ &
$\frac{\displaystyle 1563}{\displaystyle 250000}$ & 
$\frac{\displaystyle 39}{\displaystyle 50000}$ & 
$-\,\frac{\displaystyle 147}{\displaystyle 128 000}$ \\[2mm]
  & $\times \sqrt{\frac{\displaystyle m_N}{100~{\rm GeV}}}$ & 
    $\times \sqrt{\frac{\displaystyle m_N}{100~{\rm GeV}}}$ & 
  & $\times \sqrt{\frac{\displaystyle \Delta m_N}{100~{\rm GeV}}}$ &
    $\times \sqrt{\frac{\displaystyle \Delta m_N}{100~{\rm GeV}}}$ &
    $\times \sqrt{\frac{\displaystyle \Delta m_N}{100~{\rm GeV}}}$ \\[-3mm]
  & & & & & & \\
\hline
\end{tabular} }
\end{center}
\caption{\it Complete  set of the theoretical parameters  used for the
  singlet Majoron model, where  $\Delta m_N = 2(\delta {\bf m}^N)_{23}
  +     i    [(\delta     {\bf    m}^N)_{33}     -     (\delta    {\bf
  m}^N)_{22}]$.}\label{Model}
\end{table}

For our numerical analysis,  we fully specify in Table~\ref{Model} the
values  of  the theoretical  parameters  for  the  Higgs and  neutrino
sectors.   The only  parameter  that we  allow  to vary  is the  heavy
Majorana  mass scale $m_N$.   For $50~{\rm  GeV} \stackrel{<}{{}_\sim}
m_N \stackrel{<}{{}_\sim} 200~{\rm GeV}$,  the choice of parameters in
Table~\ref{Model}  leads to  an  inverted hierarchical  light-neutrino
spectrum  with  the  following  squared mass  differences  and  mixing
angles:
\begin{eqnarray}
  \label{lightspectrum}
m^2_{\nu_2}\: -\: m^2_{\nu_1} \! & = &\! (7.5\mbox{--}7.7)\times 
10^{-5}~{\rm eV}^2\; ,
\qquad
m^2_{\nu_1}\: -\: m^2_{\nu_3} \ = \ 2.44\times 10^{-3}~{\rm eV}^2\; ,
\nonumber\\
\sin^2\theta_{12} \!& = &\! 0.362\;,\quad  
\sin^2\theta_{23} \ = \ 0.341\; ,\quad \sin^2\theta_{13} \ =\ 0.047
\end{eqnarray}
and   $m_{\nu_3}  =  0$.    The  spectrum   is  compatible   with  the
light-neutrino    data    at    the   3$\sigma$    confidence    level
(CL)~\cite{JVdata}.

\begin{figure}[t]
\begin{center}
\includegraphics[scale=0.55]{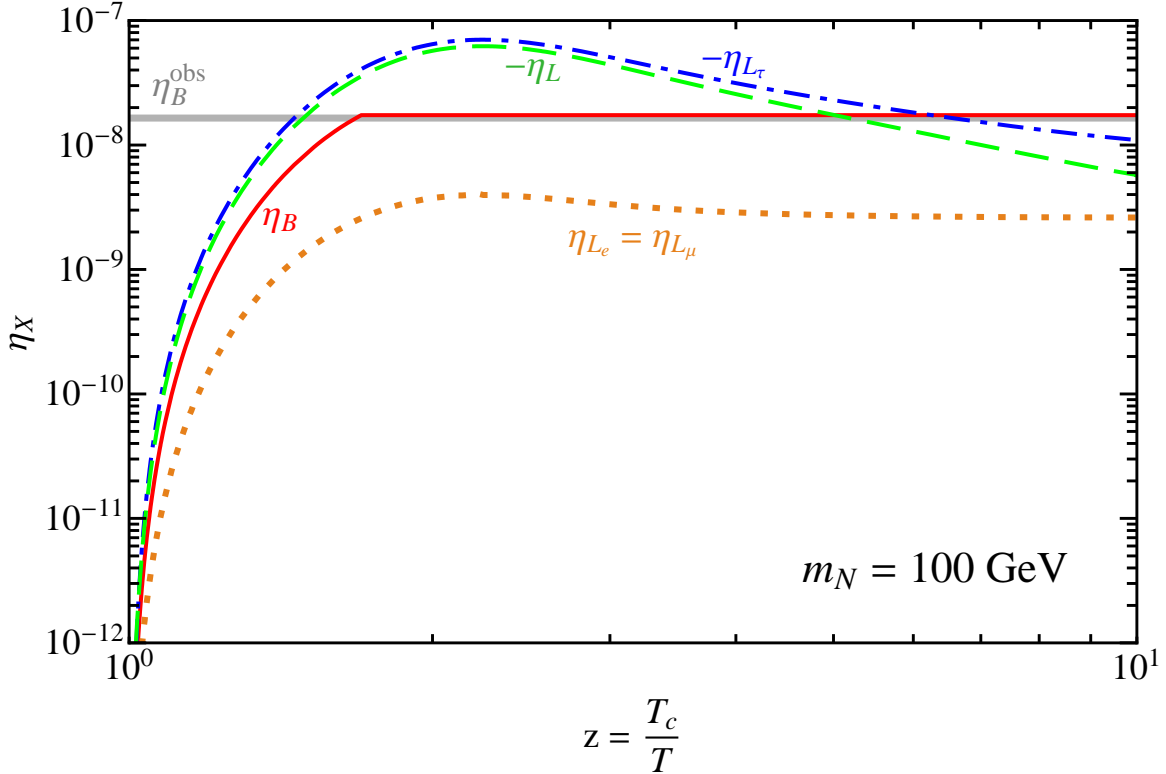}
\end{center}
\caption{\it Numerical estimates of $\eta_B$ (solid), $\eta_{L_\tau}$
  (dash-dotted), $\eta_{L_e} = \eta_{L_\mu}$ (dotted) and $\eta_L$
  (dashed) as functions of $z = T_c/T$, for a model with $m_N
  =$~100~GeV, and $\eta_{N_{\alpha}}^{\rm in} = 1$.  The model
  parameters are given in Table~\ref{Model}.
The horizontal grey line  corresponds to the observed baryon-to-photon
  ratio $\eta^{\rm  obs}_B = 1.65 \times 10^{-8}$,  after evolving the
  latter back to the higher temperature $T = T_c/10$.}
\label{fig:mN100}
\end{figure}

In  Fig.~\ref{fig:mN100}   we  present  numerical   estimates  of  the
lepton-flavour  asymmetries  $\eta_{L_{e,\mu,\tau}}$  and  the  baryon
asymmetry  $\eta_B$  as  functions  of  $z =  T_c/T$,  for  a  typical
electroweak RL  scenario with $m_N = 100$~GeV.   As initial conditions
at $T=T_c \approx 133~{\rm  GeV}$, we take $\eta_{N_{\alpha}}^{\rm in}
=  1$   for  the  heavy   neutrino  number  densities   and  vanishing
lepton-to-photon   and    baryon-to-photon   ratios,   i.e.~$\eta^{\rm
in}_{L_{e,\mu,\tau}}  = 0$  and $\eta^{\rm  in}_B =  0$.   The thermal
in-equilibrium condition $\eta_{N_{\alpha}}^{\rm in} = 1$ is expected,
since the heavy neutrinos $N_{1,2,3}$  have no chiral masses when $T >
T_c$  and  get  rapidly  thermalised by  the  sizeable  light-to-heavy
neutrino    Yukawa     couplings    $\sqrt{2}    (m_D)_{i\alpha}/    v
\stackrel{>}{{}_\sim}    10^{-7}$.     As    can    be    seen    from
Fig.~\ref{fig:mN100}, a net baryon  asymmetry $\eta_B$ is generated by
a    non-zero   $\tau$-lepton    asymmetry    $\eta_{L_\tau}$.    This
$L_\tau$-excess  is  created  before  sphalerons sharply  freeze  out,
i.e.~for  temperatures $T  \stackrel{>}{{}_\sim}  T_{\rm sph}  \approx
78$~GeV ($z \stackrel{<}{{}_\sim}  1.7$). Consequently, in the thermal
evolution  of the  Universe,  there is  a  sufficiently long  interval
$78~{\rm  GeV} \stackrel{<}{{}_\sim} T  \stackrel{<}{{}_\sim} 133~{\rm
GeV}$, where a  leptonic asymmetry can be converted  into the observed
BAU for our scenarios  with spontaneous lepton-number violation at the
electroweak scale.

\begin{figure}[t]
\begin{center}
\includegraphics[scale=0.55]{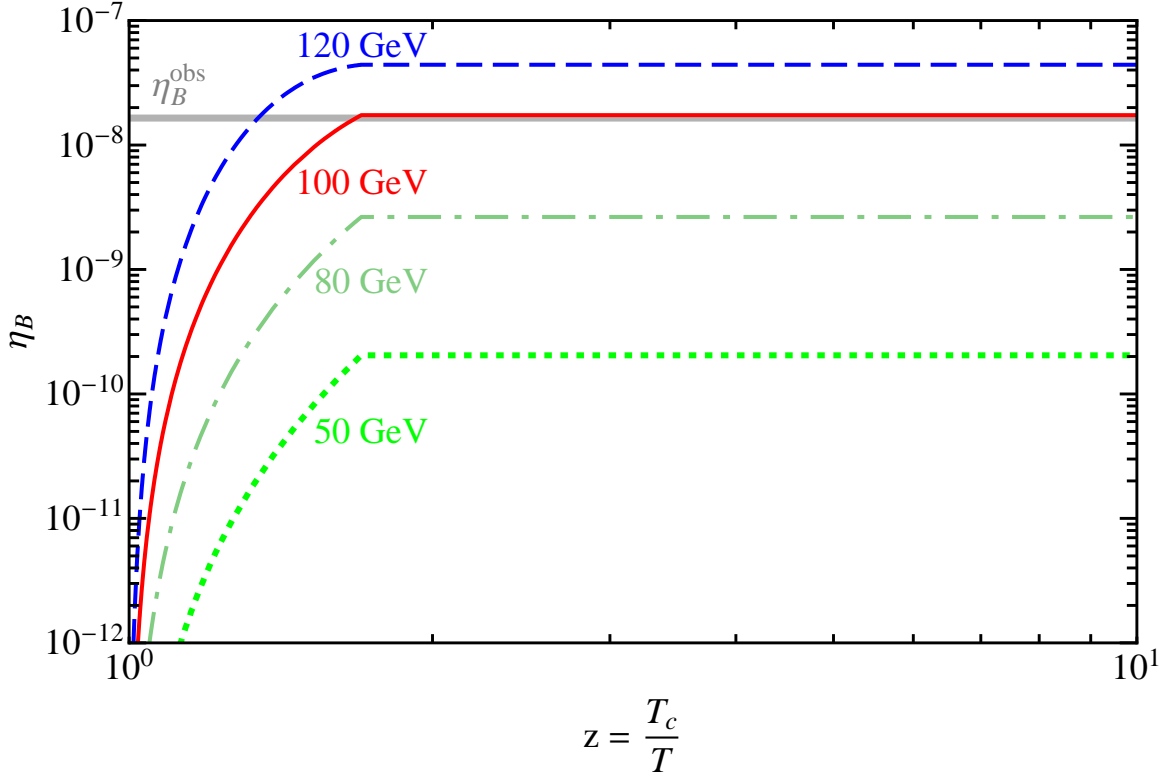}
\end{center}
\caption{\it Numerical estimates of $\eta_B$ versus $z = T_c/T$ for
$m_N = 120$~GeV (dashed), 100~GeV (solid), 80~GeV (dash-dotted),
50~GeV (dotted). The meaning of the horizontal grey line is the same
as in Fig.~\ref{fig:mN100}.}
\label{fig:mNall}
\end{figure}

Figure~\ref{fig:mNall} exhibits the dependence of the baryon-to-photon
ratio  $\eta_B$ on  $z  = T_c/T$  for  different values  of the  heavy
Majorana  mass scale  $m_N$.  We  notice  that the  lighter the  heavy
neutrinos  are, the  smaller  the created  baryon  asymmetry is.   For
example,  for heavy-neutrino masses  $m_N \sim  80$~(50)~GeV, $\eta_B$
falls short almost by one order~(two orders) of magnitude with respect
to the observed BAU $\eta^{\rm  obs}_B$.  This is a generic feature of
our electroweak  RL scenarios based  on large wash-out effects  due to
the     relatively    large     Dirac-neutrino     Yukawa    couplings
$(m_D)_{i\alpha}/v$.   If  the  heavy  neutrinos have  masses  $m_N  <
90$~GeV, their number  densities will start decreasing for  $T < m_N$,
potentially creating a net lepton asymmetry that can be converted into
$\eta^{\rm obs}_B$.  However, this  should happen above the freeze-out
temperature  $T_{\rm  sph}   \approx  78$~GeV  of  sphalerons.   Thus,
successful    electroweak   RL   requires    that   $m_N    >   T_{\rm
sph}$.\footnote{Recently, a different  leptogenesis scenario with $m_N
\ll  T_{\rm  sph}$  was  studied  in~\cite{Misha}, where  the  BAU  is
generated by sterile-neutrino  oscillations. Such a realisation relies
on  the  assumption  that  the  oscillating  sterile  neutrinos  start
evolving from a coherent state and retain their coherent nature within
the thermal plasma of the  expanding Universe.  In the singlet Majoron
model   we   have  been   studying   here   however,  $t$-channel   $2
\leftrightarrow 2$  scattering processes, such  as $JJ \leftrightarrow
\nu^C_{\alpha  R} \nu_{\alpha  R}$,  that occur  before  the EWPT  ($T
\stackrel{>}{{}_\sim}   T_c$)   are    strong,   with   rates   ${\cal
O}[\rho^4_{\alpha  \alpha}  T/(8\pi)]  \gg  H(T)$,  for  Higgs-singlet
Yukawa  couplings $\rho_{\alpha\alpha}  \sim 1$.   They  can therefore
lead to  rapid thermalization  and loss of  coherence of  the massless
right-handed  neutrinos.   Shortly  after  the  EWPT,  for  $z  =  T_c
/T\stackrel{>}{{}_\sim}   1.1$,   it   is   $\Gamma_{N_{1,2}}/H   \sim
10^9$--$10^{10}$  and $\Gamma_{N_3}/H \sim  1$--10, which  again gives
rise to  an almost  instant thermalization of  all the  heavy neutrino
mass eigenstates $N_{1,2,3}$.}

Finally,  it is  important to  comment on  the last  condition  $m_N >
T_{\rm sph}$ for scenarios with $T^l_c \neq T_c$.  This condition will
still  be  valid, as  long  as $T^l_c  >  T_{\rm  sph}$. However,  for
scenarios  with $T^l_c \stackrel{<}{{}_\sim}  T_c$, the  predicted BAU
$\eta_B$  will  sensitively depend  on  the initial  values~$\eta^{\rm
in}_{L_{e,\mu,\tau}}$ and $\eta^{\rm in}_B$ at $T = T_c$.  Instead, if
$T^l_c   \gg  T_c$   and  $m_N   \stackrel{>}{{}_\sim}   90$~GeV,  the
predictions  for  the  BAU  will  remain almost  unaffected,  even  if
$\eta^{\rm  in}_B   \sim  10^2\,  \eta^{\rm  obs}_B$  at   $T  =  10\,
T_c$~\cite{PU2}.

\setcounter{equation}{0}
\section{Astrophysical and Phenomenological Implications}\label{sec:pheno}

It is interesting  to discuss the implications of  the singlet Majoron
model for astrophysics and  low-energy phenomenology.  To quantify the
effects  of  heavy  Majorana  neutrinos,  we  define  the  new-physics
parameters
\begin{equation}
  \label{Omega}
\Omega_{ll'}\ =\ \delta_{ll'}\: -\: B^*_{lk}\,B_{l'k}\ =\
B^*_{l\alpha}\,B_{l'\alpha}\ ,
\end{equation}
where $l,\,  l' = e,\,  \mu ,\, \tau$.   Evidently, in the  absence of
light-to-heavy   neutrino  mixings,   the   parameters  $\Omega_{ll'}$
vanish. LEP and  low-energy electroweak data put severe  limits on the
diagonal parameters $\Omega_{ll}$~\cite{Ofit}:
\begin{equation}
  \label{Odiag}
\Omega_{ee}\       \leq\ 0.012\,,\qquad 
\Omega_{\mu\mu}\   \leq\ 0.0096\,,\qquad 
\Omega_{\tau\tau}\ \leq\ 0.016\, ,
\end{equation}
at  the 90\% CL.   On the  other hand,  lepton-flavour-violating (LFV)
decays, such as $\mu \to  e\gamma$~\cite{CL}, $\mu \to eee$, $\tau \to
e\gamma$,   $\tau    \to   eee$,    $\mu   \to   e$    conversion   in
nuclei~\cite{IP,LFVrev}  and  $Z  \to ll'$~\cite{KPS},  constrain  the
off-diagonal parameters $\Omega_{ll'}$,  with $l\neq l'$.  The derived
constraints   strongly   depend   on   the   heavy   neutrino   masses
$m_{N_\alpha}$  and the  size of  the Dirac  masses $(m_D)_{l\alpha}$.
However,  for models relevant  to leptogenesis,  with $(m_D)_{l\alpha}
\ll M_W$~\cite{IP}, we obtain the following limits:
\begin{equation}
  \label{Ooff}
|\Omega_{e\mu}|\    \stackrel{<}{{}_\sim}\ 0.0001\,,\qquad 
|\Omega_{e\tau}|\   \stackrel{<}{{}_\sim}\ 0.02\,,\qquad 
|\Omega_{\mu\tau}|\ \stackrel{<}{{}_\sim}\ 0.02\, ,
\end{equation}
including the recent BaBar data  on LFV $\tau$ decays~\cite{Babar}. 

The predictions for LFV decays  in models of resonant leptogenesis has
been extensively  discussed in~\cite{PU2}. Since  our results obtained
in Section~\ref{sec:EWPT}  agree well  with this earlier  analysis, we
will  not  repeat  the details  of  this  study  here. Here,  we  only
reiterate the  fact that successful electroweak RL  requires that $m_N
\stackrel{>}{{}_\sim}  100~{\rm GeV}$.   This latter  constraint gives
rise to the following upper limits:
\begin{equation}
  \label{Olepto}
\Omega_{ee}\ \stackrel{<}{{}_\sim}\ 2.2\times 10^{-4}\;,\qquad
|\Omega_{e\mu}|\ \stackrel{<}{{}_\sim}\ 8.3\times 10^{-5}\; ,\qquad
\Omega_{\mu\mu}\ \stackrel{<}{{}_\sim}\ 3.1\times 10^{-5}\; ,
\end{equation}
whereas   all   remaining   parameters   $\Omega_{ll'}$   are   ${\cal
O}(10^{-8})$ and so unobservably  small.  All these limits are deduced
by using the model parameters of~Table~\ref{Model}.

%******************************************************************
%%% Figure 2 on  J - l - l' and   J - q - q   couplings
%******************************************************************
\begin{figure}

\begin{center}
\begin{picture}(360,100)(0,0)
\SetWidth{0.8}

\ArrowLine(0,60)(20,60)\ArrowLine(60,60)(80,60)
\Photon(20,60)(60,60){2}{4}\Text(40,67)[b]{$W^-$}
\ArrowLine(20,60)(40,40)\ArrowLine(40,40)(60,60)
\Text(30,40)[r]{$N_\alpha$}\Text(55,40)[l]{$N_\beta$}
\DashLine(40,40)(40,20){3}
\Text(0,65)[b]{$l$}\Text(80,65)[b]{$l'$}
\Text(45,20)[l]{$J$}
\Text(40,0)[]{\bf (a)}

\ArrowLine(140,60)(160,60)\ArrowLine(200,60)(220,60)
\DashArrowLine(160,60)(200,60){3}\Text(180,67)[b]{$G^-$}
\ArrowLine(160,60)(180,40)\ArrowLine(180,40)(200,60)
\Text(170,40)[r]{$N_\alpha$}\Text(195,40)[l]{$N_\beta$}
\DashLine(180,40)(180,20){3}
\Text(140,65)[b]{$l$}\Text(220,65)[b]{$l'$}
\Text(185,20)[l]{$J$}
\Text(180,0)[]{\bf (b)}

\ArrowLine(260,80)(300,80)\ArrowLine(300,80)(340,80)
\Photon(300,80)(300,60){2}{2}\Text(306,70)[l]{$Z$}
\ArrowArc(300,50)(10,90,270)\ArrowArc(300,50)(10,-90,90)
\Text(288,50)[r]{$N_\alpha$}\Text(315,50)[l]{$N_\beta$}
\DashLine(300,40)(300,20){3}
\Text(260,85)[b]{$l,q$}\Text(340,85)[b]{$l,q$}
\Text(305,20)[l]{$J$}
\Text(300,0)[]{\bf (c)}

\end{picture}\\[0.7cm]
\end{center}

\caption{\it Loop-induced couplings of the Majoron to charged leptons $l,\,
  l'$ and quarks $q$.}\label{fig:J}

\end{figure}
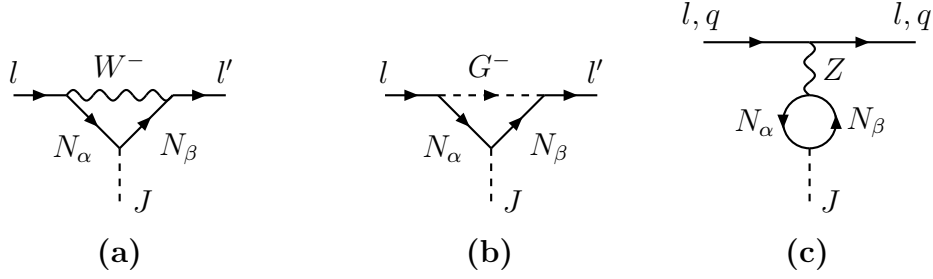

In the  singlet Majoron  model under study,  there are  additional LFV
decays  for the  muon and  the  tau-lepton that  involve the  Majoron,
i.e.~$\mu \to J e$, $\tau \to J  e$ and $\tau \to J \mu$.  As shown in
Fig.~\ref{fig:J},  these  LFV decays  are  induced  by heavy  Majorana
neutrinos at the 1-loop  level.  Detailed analytic expressions for the
loop-induced couplings $Jll'$ and $Jqq$,  where $q$ is a quark, may be
found  in~\cite{APMaj}.   To  leading  order  in  $\Omega_{ll'}$,  the
prediction for the LFV decay $l^- \to l'^-J$ is
\begin{equation}
R (l \to l' J)\ \equiv\ 
\frac{\Gamma (l^- \to l'^- J)}{\Gamma (l^- \to l'^- \nu_l \bar{\nu}_{l'})}\ 
=\ \frac{3\alpha_w}{8\pi}\ t^2_\beta\, |\Omega_{ll'}|^2\;
\frac{M^2_W}{m^2_l}\; \frac{\lambda^4_N}{(1 - \lambda_N )^2}\;
\Bigg( 1\: +\: \frac{\ln \lambda_N}{1-\lambda_N}\Bigg)^2\; , 
\end{equation}
where $\lambda_N  = m^2_N/M^2_W$. For $\lambda_N =  1$, the prediction
for the observable $R (l \to l' J)$ takes on the simpler form:
\begin{equation}
  \label{Robs}
R (l \to l' J)\ =\ \frac{3\alpha_w}{32\pi}\ t^2_\beta\, |\Omega_{ll'}|^2\;
\frac{M^2_W}{m^2_l}\; .
\end{equation}
The   requirement    for   successful   electroweak    RL,   i.e.~$m_N
\stackrel{>}{{}_\sim}  100$~GeV, gets  translated  into the  following
upper bounds:
\begin{equation}
  \label{Rtheory}
R ( \mu \to eJ)\ \stackrel{<}{{}_\sim}\ 2.7\times 10^{-6}\;,\quad
R ( \tau \to eJ)\ \stackrel{<}{{}_\sim}\ 4.6\times 10^{-14}\;,\quad
R ( \tau \to \mu J)\ \stackrel{<}{{}_\sim}\ 6.7\times 10^{-15}\;.
\end{equation}
On the experimental side, however, the following upper limits are quoted:
\begin{eqnarray}
R(\mu \to e J) \!&\leq&\! 2.6\times 10^{-6}\,,\quad 
                                \mbox{at 90\%  CL~\cite{AJexp};}\nonumber\\
R(\tau \to e J)\!&\leq&\! 1.5\times 10^{-2}\,,\quad 
                                \mbox{at 95\% CL~\cite{ARGUS};}\\
R(\tau \to \mu J)\! &\leq &\! 2.6\times 10^{-2}\,,\quad 
                                \mbox{at 95\% CL~\cite{ARGUS}}.\nonumber 
\end{eqnarray}
It is interesting to remark that  the predicted value for $R ( \mu \to
eJ)$  is close to  the present  experimental sensitivity,  whereas the
other decay  modes turn  out to  be very suppressed  for the  given RL
model with  inverted light-neutrino hierarchy.  Had we  chosen a model
with normal hierarchy,  the decay rates $R(\tau \to  e J)$ and $R(\tau
\to \mu J)$ would have been enhanced by a factor $\sim 10^8$, but they
will still  be rather  small ${\cal O}(10^{-6})$  to be  observed; the
predictions  generally lie  4 orders  of magnitude  below  the current
experimental upper bounds.

Useful constraints on  the parameters of the theory  are obtained from
astrophysics as well~\cite{GGR}.  Specifically, observational evidence
of cooling rates  of white dwarfs implies that  the interaction of the
Majoron  to electrons,  $g_{Jee}  J \bar{e}  i\gamma_5  e$, should  be
sufficiently weak and the coupling $g_{Jee}$ must obey the approximate
upper bound~\cite{Astro}:
\begin{equation}
  \label{gJee}
|g_{Jee}|\ \stackrel{<}{{}_\sim}\ 10^{-12}\ .
\end{equation}
The  above limit gets  further consolidated  by considerations  of the
helium ignition process in red  giants, leading to the excluded range:
$3\times      10^{-13}\,      \stackrel{<}{{}_\sim}\,      |g_{Jee}|\,
\stackrel{<}{{}_\sim}\,  6\times   10^{-7}$.   To  leading   order  in
$\Omega_{ll}$,   the  loop-induced   coupling   $g_{Jee}$,  is   given
by~\cite{APMaj}:
\begin{equation}
  \label{Jee}
g_{Jee}\ =\ \frac{g_w\,\alpha_w}{16\pi}\ \frac{m_e}{M_W}\  t_\beta\
\lambda_N\, 
\Bigg[\, \Omega_{ee}\ \frac{\lambda_N}{1\, -\, \lambda_N}\
\Bigg(\, 1\: +\: \frac{\ln\lambda_N}{1\, -\, \lambda_N}\,\Bigg)\: +\:
\frac{1}{2}\; \sum\limits_{l=e,\mu,\tau} \Omega_{ll}\, \Bigg]\;.
\end{equation}
If  $\lambda_N\gg  1$,  the  expression  for  the  coupling  $g_{Jee}$
simplifies to
\begin{equation}
  \label{approxJee}
g_{Jee}\ =\ \frac{g_w\,\alpha_w}{32\pi}\ \frac{m_e}{M_W}\  t_\beta\
\lambda_N\, 
\Big(\, \Omega_{\mu\mu}\: +\: \Omega_{\tau\tau}\: -\:
\Omega_{ee}\,\Big)\; ,
\end{equation}
whilst for $\lambda_N = 1$ $g_{Jee}$ becomes
\begin{equation}
  \label{Jeelambda}
g_{Jee}\ =\ \frac{g_w\,\alpha_w}{32\pi}\ \frac{m_e}{M_W}\  t_\beta\
\Big(\, \Omega_{\mu\mu}\: +\: \Omega_{\tau\tau}\,\Big)\; .
\end{equation}
Given the limits~(\ref{Olepto}) for successful RL, we can estimate that
\begin{equation}
g_{Jee}\ \stackrel{<}{{}_\sim}\ -3.3\times 10^{-17}\; ,
\end{equation}
which   passes   comfortably   the  astrophysical   constraint   given
in~(\ref{gJee}).

Useful   astrophysical   constraints  may   also   be  obtained   from
considerations  of  the  cooling  rate  of  neutron  stars~\cite{GGR}.
Neutron  stars  will loose  energy  by  Majoron  emission through  the
interaction: $g_{J{\cal N}{\cal  N}}\, J\; \overline{\cal N} i\gamma_5
{\cal N}$, where ${\cal N}$ is a nucleon, specifically a neutron.  The
observational limit on $g_{J{\cal N}{\cal N}}$ is~\cite{NI}
\begin{equation}
  \label{JNN}
g_{J{\cal N}{\cal  N}}\ \stackrel{<}{{}_\sim}\ 10^{-9}\ . 
\end{equation}
On the  other hand,  the theoretical prediction  for $g_{Jqq}$  at the
quark level is
\begin{equation}
  \label{Jqq}
g_{Jqq}\ =\ \frac{g_w\,\alpha_w}{32\pi}\ \frac{m_q}{M_W}\  t_\beta\
\lambda_N\, \Big(\, \Omega_{ee}\: +\: \Omega_{\mu\mu}\: +\:
\Omega_{\tau\tau}\,\Big) \; .
\end{equation}
{}From  naive   dimensional  analysis  arguments,   one  expects  that
$g_{J{\cal N}{\cal  N}} \sim (m_{\cal N}/m_q) g_{Jqq}$.   In this way,
one may estimate that
\begin{equation}
g_{J{\cal N}{\cal N}} \approx 7\times 10^{-10}\; , 
\end{equation}
after taking into consideration the limits stated in~(\ref{Olepto}).

Cosmic microwave background (CMB) data and BBN put stringent limits on
the  maximum  number  of  weakly-interacting relativistic  degrees  of
freedom,  such as  light neutrinos  and  Majorons~\cite{BKLMS,IST}. In
particular,  the  allowed  range  obtained for  the  effective  number
$N_\nu$    of    left-handed   neutrino    species    is   $N_\nu    =
2.70^{+0.91}_{-1.32}$ at  the 68\%  CL~\cite{IST}. The upper  bound on
$N_\nu$ may naively be translated into an upper limit on $\Delta N_\nu
= N_\nu -  3 = 0.61$ of extra effective neutrino  species beyond the 3
SM  left-handed neutrinos.   The singlet  Majoron  contributes $\Delta
N_\nu  =  (\frac{1}2\times  \frac{8}7)^{4/3}  \approx 0.474$,  if  its
freeze-out   or  decoupling   temperature  $T_J$   is  equal   to  the
corresponding one $T_\nu$ of the neutrinos.  Although this result does
not pose by itself a  serious limitation on the singlet Majoron model,
it can be  estimated, however, that $T_J \gg  T_\nu \approx 1$~MeV and
the  contribution   of  $J$  to  $\Delta  N_\nu$   becomes  even  more
suppressed.   Specifically,   the  freeze-out  temperature   $T_J$  is
determined when the annihilation  rate of Majorons through the process
$JJ \to \nu\nu$ becomes smaller  than the Hubble expansion rate $H(T)$
of the Universe.  The annihilation process $JJ \to \nu\nu$ is mediated
by  the  $H$ and  $S$  bosons  in the  $s$-channel  and  by the  heavy
neutrinos  $N_{1,2,3}$  in the  $t$-channel.   Considering the  latter
reactions only, one may naively estimate that
\begin{equation}
  \label{TJ}
\frac{T_J}{T_\nu}\ \sim\ \Bigg(\frac{G^2_F\, m^4_N}{\Omega_{ee}^2\, 
 t^4_\beta}\Bigg)^{1/3}\ \sim\ 10^2\, \mbox{--}\, 10^3\; .
\end{equation}
A  similar  value  for  $T_J/T_\nu$  is obtained  if  the  $S,H$-boson
exchange   processes   are   used   for  the   model   parameters   of
Table~\ref{Model}.  Thus, the freeze-out temperature $T_J$ lies in the
range $0.1$--1~GeV, namely about the quark-hadron deconfinement phase.
In this  epoch of the  Universe, the effective number  of relativistic
degrees  of  freedom  is  $g_*(T_J)  \approx  66$.  Then,  the  actual
contribution of the Majoron to  $\Delta N_\nu$ is reduced with respect
to  the $T_J =  T_\nu$ case  by a  factor $(g_*(T_\nu)/g_*(T_J))^{4/3}
\approx 0.016$ to the value $\Delta N_\nu \approx 0.008$, which is far
below the present and future observational sensitivity~\cite{IST}.

Finally, singlet  Majorons $J$ and  singlet scalars $S$ may  also give
rise  to  interesting  collider phenomenology~\cite{ASJ}  through  the
singlet-doublet    mixing   parameter    $\delta$   in    the   scalar
potential~(\ref{LV}).    However,  since  $\delta   \ll  \lambda_\Phi$
(cf.~Table~\ref{Model}),  the  singlet  Majoron scenario  under  study
predicts  a rather small  mixing angle  $s_\theta \approx  -0.1$.  The
production cross section  of $S$, via the process  $e^+e^-\to Z S$, is
then suppressed  with respect  to the SM  one by a  factor $s^2_\theta
\approx  0.01$. Moreover,  the  so-produced Higgs  singlets may  decay
quasi-invisibly into a pair of  Majorons $J$, which makes difficult to
fully rule  out such a  scenario by LEP2  data or at the  LHC.  Future
high-energy  $e^+e^-$  colliders of  higher  luminosity will  severely
constrain the allowed parameter space of this singlet Majoron model.

\setcounter{equation}{0}
\section{Conclusions}\label{sec:concl}

The origin  of CP  violation in nature  still remains an  open physics
question.  If  CP violation  originates from the  SSB of the  SM gauge
group, the original  scenario~\cite{FY} of GUT-scale leptogenesis will
be excluded.  Similar will be  the fate of all high-scale leptogenesis
models, if the source of lepton-number  violation is due to the SSB of
a global U(1)$_l$ symmetry at  the electroweak scale. In this paper we
have  shown  how  resonant  leptogenesis  at the  EWPT  constitutes  a
realistic  alternative  for  successful  baryogenesis in  models  with
spontaneous lepton-number violation.  Specifically, we have considered
a  minimal extension  of  the  SM, the  singlet  Majoron model,  which
includes  right-handed  neutrinos and  a  complex  singlet field  that
carries a non-zero lepton number.  Depending on the form of the scalar
potential, the lepton number  can get broken spontaneously through the
VEV of the SM Higgs  doublet.  Taking into consideration the Boltzmann
dynamics of sphaleron effects, we  have analysed the BAU for different
values  of the  Majorana  mass scale  $m_N$  within the  context of  a
benchmark   scenario    whose   model   parameters    are   given   in
Table~\ref{Model}.   The  generic  constraint from  having  successful
electroweak RL is that  $m_N \stackrel{>}{{}_\sim} T_{\rm sph}$, where
$T_{\rm  sph} \approx  78$~GeV is  the freeze-out  temperature  of the
sphalerons.

The singlet Majoron model  predicts a massless Goldstone particle, the
Majoron $J$.  The Majoron can be produced via the LFV decays, $\mu \to
J  e$,  $\tau  \to  J  \mu$  and $\tau  \to  J  e$.   Considering  the
constraints  from  successful  electroweak  RL and  the  astrophysical
limits derived from  the cooling rate of neutron  stars, we have found
that the decay mode $\mu \to  J e$ is the most promising channel, with
sizeable branching fraction  that can be looked for  in the next-round
low-energy experiments.

The predictions obtained for the BAU  in this study are limited by the
approximations   that  are   inherent  in   the  calculation   of  the
non-perturbative  sphaleron dynamics. The  predicted values  should be
regarded as order-of-magnitude  estimates, since the $(B+L)$-violating
sphaleron transitions crucially depend  on the parameter $\kappa$ that
varies by a  factor of 10 or so.  It would  therefore be very valuable
to  go  beyond  the  current  approximation methods  and  improve  the
computation  of  the out-of-equilibrium  sphaleron  dynamics during  a
second-order electroweak phase transition.

\bigskip

\subsection*{Acknowledgements}
I  thank   Roger  Barlow,  George  Lafferty  and   Olga  Igonkina  for
discussions  concerning the  current status  of Majoron  searches, and
Frank Deppisch  for a critical  reading of the manuscript.   This work
was supported in part by the STFC research grant: PP/D000157/1.

\newpage

\def\theequation{\Alph{section}.\arabic{equation}}
%\begin{appendix}

%\setcounter{equation}{0}
%\section{Majoron-Induced LFV Decays}

%\end{appendix}


\newpage
\begin{thebibliography}{99}


\bibitem{FY} M. Fukugita and T. Yanagida, Phys.\ Lett.\ B {\bf 174}
(1986) 45.

\bibitem{WMAP} J.~Dunkley {\it et al.}  [WMAP Collaboration],
  %``Five-Year Wilkinson Microwave Anisotropy Probe (WMAP) Observations:
  %Likelihoods and Parameters from the WMAP data,''
  arXiv:0803.0586 [astro-ph].

\bibitem{reviews}    For  recent    reviews,   see,\\ 
W.~Buchmuller, R.~D.~Peccei and T.~Yanagida,
  %``Leptogenesis as the origin of matter,''
  Ann.\ Rev.\ Nucl.\ Part.\ Sci.\  {\bf 55} (2005) 311
  [arXiv:hep-ph/0502169];\\
S.~Davidson, E.~Nardi and Y.~Nir,
%``Leptogenesis,''
arXiv:0802.2962 [hep-ph].

\bibitem{KRS} V.~A.~Kuzmin, V.~A.~Rubakov and M.~E.~Shaposhnikov,
  Phys.\ Lett.\ B {\bf 155} (1985) 36.

\bibitem{seesaw}  P.~Minkowski, Phys.\ Lett.\ B {\bf 67} (1977) 421;\\
M. Gell-Mann, P.  Ramond and R. Slansky, in  {\em Supergravity}, 
eds.~D.Z.  Freedman  and  P.~van Nieuwenhuizen  (North-Holland,  Amsterdam,
1979);\\ 
T.  Yanagida, in  Proc.\ of  the {\em  Workshop on  the Unified
Theory and the  Baryon Number in the Universe},  Tsukuba, Japan, 1979,
eds.\ O.~Sawada and  A.~Sugamoto;\\ 
R.~N.~Mohapatra and G.~Senjanovi\'c, Phys.\ Rev.\ Lett.\ {\bf 44} (1980) 912.

\bibitem{gravitino} M.~Y.~Khlopov and A.~D.~Linde,
  %``Is It Easy To Save The Gravitino?,''
  Phys.\ Lett.\  B {\bf 138} (1984) 265;\\
J.~R.~Ellis, J.~E.~Kim and D.~V.~Nanopoulos,
  %``Cosmological Gravitino Regeneration And Decay,''
  Phys.\ Lett.\  B {\bf 145} (1984) 181;\\
J.~R.~Ellis, D.~V.~Nanopoulos and S.~Sarkar,
  %``The Cosmology Of Decaying Gravitinos,''
  Nucl.\ Phys.\ B {\bf 259} (1985) 175;\\
J.~R.~Ellis, G.~B.~Gelmini, J.~L.~Lopez, D.~V.~Nanopoulos and S.~Sarkar,
%``Astrophysical Constraints On Massive Unstable Neutral Relic Particles,''
  Nucl.\ Phys.\ B {\bf 373} (1992) 399;\\
 R.~H.~Cyburt, J.~R.~Ellis, B.~D.~Fields and K.~A.~Olive,
  Phys.\ Rev.\ D {\bf 67} (2003) 103521;\\
  J.~R.~Ellis, K.~A.~Olive and E.~Vangioni, Phys.\ Lett.\ B {\bf 619}
 (2005) 30;\\ 
 M.~Kawasaki, K.~Kohri and T.~Moroi,  Phys.\ Lett.\ B {\bf
  625} (2005) 7;    Phys.\ Rev.\ D {\bf 71}
  (2005) 083502.

\bibitem{APRD} A.~Pilaftsis, Phys.\ Rev.\ D {\bf 56} (1997) 5431;
 Int.\ J. Mod.\ Phys.\ A {\bf 14} (1999) 1811.

\bibitem{LiuSegre} 
J.~Liu and G.~Segr\'e, Phys.\ Rev.\ D {\bf 48} (1993) 4609;\\
M.~Flanz, E.A.~Paschos and U.~Sarkar, Phys.\ Lett.\ B~{\bf 345} (1995) 248;\\ 
L.~Covi, E.~Roulet and F.~Vissani, Phys.\ Lett.\  B~{\bf 384} (1996) 169.

\bibitem{PU} A.~Pilaftsis and T.~E.~J. Underwood, Nucl.\ Phys.\ B {\bf 692}
    (2004) 303.

\bibitem{EMX} T.~Endoh, T.~Morozumi and Z.~h.~Xiong,
  %``Primordial lepton family asymmetries in seesaw model,''
  Prog.\ Theor.\ Phys.\  {\bf 111} (2004) 123.

\bibitem{APtau} A. Pilaftsis, Phys.\ Rev.\ Lett.\ {\bf 95} (2005)
  081602.

\bibitem{PU2} A.~Pilaftsis and T.~E.~J. Underwood, Phys.\ Rev.\ D~{\bf
  72} (2005) 113001.

\bibitem{APZPC} A. Pilaftsis, Z.\ Phys.\ C {\bf 55} (1992) 275.

\bibitem{DGP} A.~Datta, M.~Guchait and A.~Pilaftsis,
  %``Probing lepton number violation via majorana neutrinos at hadron
  %supercolliders,''
  Phys.\ Rev.\  D {\bf 50} (1994) 3195;\\
T.~Han and B.~Zhang,
  %``Signatures for Majorana neutrinos at hadron colliders,''
  Phys.\ Rev.\ Lett.\  {\bf 97} (2006) 171804;\\
F.~del Aguila, J.~A.~Aguilar-Saavedra and R.~Pittau,
  %``Heavy neutrino signals at large hadron colliders,''
  JHEP {\bf 0710} (2007) 047;\\
S.~Bray, J.~S.~Lee and A.~Pilaftsis,
  %``Resonant CP violation due to heavy neutrinos at the LHC,''
  Nucl.\ Phys.\  B {\bf 786} (2007) 95;\\
S.~Bar-Shalom, G.~Eilam, T.~Han and A.~Soni,
%``Charged Higgs Boson Effects in the Production and Decay of a Heavy Majorana
%Neutrino at the LHC,''
  arXiv:0803.2835 [hep-ph].

\bibitem{RLpapers} 
  T.~Hambye, J.~March-Russell and S.~M.~West,
%``TeV scale resonant leptogenesis from supersymmetry breaking,''
  JHEP {\bf 0407} (2004) 070;\\
C.~H.~Albright and S.~M.~Barr,
  %``Resonant leptogenesis in a predictive SO(10) grand unified model,''
  Phys.\ Rev.\  D {\bf 70} (2004) 033013;\\
E.~K.~Akhmedov, M.~Frigerio and A.~Y.~Smirnov,
  %``Probing the seesaw mechanism with neutrino data and leptogenesis,''
  JHEP {\bf 0309} (2003) 021;\\
  E.~J.~Chun,
%``TeV leptogenesis in Z-prime models and its collider probe,''
  Phys.\ Rev.\ D {\bf 72} (2005) 095010;\\
S.~M.~West,
  %``Neutrino Masses And Tev Scale Resonant Leptogenesis From Supersymmetry
  %Breaking,''
  Mod.\ Phys.\ Lett.\  A {\bf 21} (2006) 1629;\\
Z.~z.~Xing and S.~Zhou,
  %``Tri-bimaximal Neutrino Mixing and Flavor-dependent Resonant
  %Leptogenesis,'' 
  Phys.\ Lett.\  B {\bf 653} (2007) 278;\\
K.~S.~Babu, A.~G.~Bachri and Z.~Tavartkiladze,
  %``Predictive Model of Inverted Neutrino Mass Hierarchy and Resonant
  %Leptogenesis,''
  arXiv:0705.4419 [hep-ph];\\
S.~Uhlig,
  %``Minimal lepton flavour violation and leptogenesis with exclusively
  %low-energy CP violation,''
  JHEP {\bf 0711} (2007) 066;\\
V.~Cirigliano, A.~De Simone, G.~Isidori, I.~Masina and A.~Riotto,
  %``Quantum Resonant Leptogenesis and Minimal Lepton Flavour Violation,''
  JCAP {\bf 0801} (2008) 004.

\bibitem{RLextra}  For studies of  RL in  higher-dimensional theories,
see\\ 
A. Pilaftsis, 
%``Leptogenesis in theories with large extra dimensions,''
Phys.\ Rev.\  D {\bf 60} (1999) 105023;\\ 
A.~D.~Medina and C.~E.~M.~Wagner,
%``Soft leptogenesis in warped extra dimensions,''
JHEP {\bf 0612} (2006) 037;\\ 
T.~Gherghetta, K.~Kadota and M.~Yamaguchi,
%``Warped Leptogenesis with Dirac Neutrino Masses,''
Phys.\ Rev.\  D {\bf 76} (2007) 023516;
M.~T.~Eisele,  %``Leptogenesis With Many Neutrinos,''
Phys.\ Rev.\  D {\bf 77} (2008) 043510.

\bibitem{soft} Y.~Grossman, T.~Kashti, Y.~Nir and E.~Roulet,
  %``Leptogenesis from supersymmetry breaking,''
  Phys.\ Rev.\ Lett.\  {\bf 91} (2003) 251801;\\
G.~D'Ambrosio, G.~F.~Giudice and M.~Raidal,
  %``Soft leptogenesis,''
  Phys.\ Lett.\  B {\bf 575} (2003) 75;\\
C.~S.~Fong and M.~C.~Gonzalez-Garcia,
  %``Flavoured Soft Leptogenesis,''
  arXiv:0804.4471 [hep-ph].

\bibitem{rad} R.~Gonzalez Felipe, F.~R.~Joaquim and B.~M.~Nobre,
  %``Radiatively induced leptogenesis in a minimal seesaw model,''
  Phys.\ Rev.\  D {\bf 70} (2004) 085009;\\
K. Turzynski, Phys.\ Lett.\ B~{\bf 589} (2004) 135;\\
G.~C.~Branco, R.~Gonzalez Felipe, F.~R.~Joaquim and B.~M.~Nobre,
  %``Enlarging the window for radiative leptogenesis,''
  Phys.\ Lett.\  B {\bf 633} (2006) 336;\\
G.~C.~Branco, A.~J.~Buras, S.~Jager, S.~Uhlig and A.~Weiler,
  %``Another look at minimal lepton flavour violation, l(i) --> l(j) gamma,
  %leptogenesis, and the ratio M(nu)/Lambda(LFV),''
  JHEP {\bf 0709} (2007) 004.

\bibitem{HR} P.~Hernandez and N.~Rius,
  %``Neutral heavy leptons and electroweak baryogenesis,''
  Nucl.\ Phys.\  B {\bf 495} (1997) 57.

\bibitem{CMP} 
Y.~Chikashige, R.~N.~Mohapatra and R.~D.~Peccei,
%``Are There Real Goldstone Bosons Associated With Broken Lepton Number?,''
Phys.\ Lett.\  B {\bf 98} (1981) 265;\\
J. Schechter and J.~W.~F. Valle, Phys.\ Rev.\ D~{\bf 25} (1982) 774.

\bibitem{APMaj} A. Pilaftsis, Phys.\ Rev.\ D~{\bf 49} (1994) 2398.

\bibitem{LEPHiggs} S.~Schael {\it et al.}  [ALEPH Collaboration],
  %``Search for neutral MSSM Higgs bosons at LEP,''
  Eur.\ Phys.\ J.\  C {\bf 47} (2006) 547.

\bibitem{EWSM} M.~B.~Gavela, P.~Hernandez, J.~Orloff, O.~Pene and C.~Quimbay,
  %``Standard model CP violation and baryon asymmetry. Part 2: Finite
  %temperature,''
  Nucl.\ Phys.\  B {\bf 430} (1994) 382;\\
V.~A.~Rubakov and M.~E.~Shaposhnikov,
  %``Electroweak baryon number non-conservation in the early universe and in
  %high-energy collisions,''
  Usp.\ Fiz.\ Nauk {\bf 166} (1996) 493
  [Phys.\ Usp.\  {\bf 39} (1996) 461].


\bibitem{PP} J. Papavassiliou and A. Pilaftsis, Phys.\ Rev.\
  Lett.\ {\bf 75} (1995) 3060; Phys.\ Rev.\ D {\bf 53}
  (1996) 2128; Phys.\ Rev.\ D {\bf 54} (1996) 5315.

\bibitem{APNPB} A. Pilaftsis, Nucl.\ Phys.\ B {\bf 504} (1997) 61.

\bibitem{PT} 
  J.~M.~Cornwall and J.~Papavassiliou,
  %``Gauge Invariant Three Gluon Vertex in QCD,''
  Phys.\ Rev.\  D {\bf 40} (1989) 3474;\\
  %%CITATION = PHRVA,D40,3474;%% 
J.~Papavassiliou,
  %``Gauge Invariant Proper Selfenergies And Vertices In Gauge Theories With
  %Broken Symmetry,''
  Phys.\ Rev.\  D {\bf 41} (1990) 3179;\\
 D.~Binosi and J.~Papavassiliou,
  %``The pinch technique to all orders,''
  Phys.\ Rev.\  D {\bf 66} (2002) 111901;
%  [arXiv:hep-ph/0208189];
  %%CITATION = PHRVA,D66,111901;%%
%D.~Binosi and J.~Papavassiliou,
  %``Pinch technique self-energies and vertices to all orders in perturbation
  %theory,''
  J.\ Phys.\ G {\bf 30} (2004)~203.
 % [arXiv:hep-ph/0301096].

\bibitem{JKapusta} For example, see,\\
J.~I. Kapusta, {\it Finite-Temperature Field Theory}, (Cambridge
University Press, Cambridge, 1989).

\bibitem{KPSmatrix} J.~G. K\"orner, A. Pilaftsis and K. Schilcher, Phys.\
Rev.\ D~{\bf 47} (1993) 1080.

\bibitem{Minkowski} P. Minkowski, preprint BUTP-94/10 (unpublished).

\bibitem{AS} {\it Handbook of Mathematical Functions}, edited by
  M. Abramowitz and I.~A. Stegun (Verlag Harri Deutsch, Frankfurt,
  1984).

\bibitem{MEC} M.~E.~Carrington,
  %``The Effective potential at finite temperature in the Standard Model,''
  Phys.\ Rev.\  D {\bf 45} (1992) 2933;\\
J.~M. Cline, K. Kainulainen and K.~A. Olive, Phys.\ 
  Rev.\ D {\bf 49} (1994) 6394.

\bibitem{AM} P.~Arnold and L.~D.~McLerran, Phys.\ Rev.\ D {\bf 36} (1987) 581.

\bibitem{CLMW} L.~Carson, X.~Li, L.~D.~McLerran and R.~T.~Wang, Phys.\
  Rev.\ D {\bf 42} (1990) 2127.

\bibitem{KS} S.~Y.~Khlebnikov and M.~E.~Shaposhnikov, Nucl.\ Phys.\ B
  {\bf 308} (1988) 885;\\
J.~A.~Harvey and M.~S.~Turner, Phys.\ Rev.\ D {\bf 42}
  (1990) 3344.

\bibitem{LS} M.~Laine and M.~E.~Shaposhnikov, Phys.\ Rev.\ D {\bf 61}
  (2000) 117302.

\bibitem{Kersten/Smirnov} J.~Kersten and A.~Y.~Smirnov,
  %``Right-Handed Neutrinos at LHC and the Mechanism of Neutrino Mass
  %Generation,''
  Phys.\ Rev.\  D {\bf 76} (2007) 073005.

\bibitem{JVdata} For an updated analysis, see, M.~Maltoni, T.~Schwetz,
 M.~A.~Tortola and J.~W.~Valle, New J. Phys.\ {\bf 6} (2004) 122.

\bibitem{Misha}  M.~E.~Shaposhnikov, arXiv:0804.4542 [hep-ph];\\
For earlier considerations, see,\\
E.~K.~Akhmedov, V.~A.~Rubakov and A.~Y.~Smirnov,
  %``Baryogenesis via neutrino oscillations,''
  Phys.\ Rev.\ Lett.\  {\bf 81} (1998) 1359;\\
 T.~Asaka and M.~Shaposhnikov,
  %``The nuMSM, dark matter and baryon asymmetry of the universe,''
  Phys.\ Lett.\  B {\bf 620} (2005) 17.

\bibitem{Ofit} C.~P.~Burgess, S.~Godfrey, H.~Konig, D.~London and
  I.~Maksymyk, Phys.\ Rev.\ D {\bf 49} (1994) 6115;\\ 
  E.~Nardi, E.~Roulet and D.~Tommasini, Phys.\ Lett.\ B~{\bf 327} (1994) 319;\\
  S.~Bergmann and A.~Kagan, Nucl.\ Phys.\ B~{\bf 538} (1999) 368;\\
  F.~del Aguila, J.~de Blas and M.~Perez-Victoria,
  %``Effects of new leptons in Electroweak Precision Data,''
  arXiv:0803.4008 [hep-ph].

\bibitem{CL} T.~P.~Cheng and L.~F.~Li, Phys.\ Rev.\ Lett.\ {\bf 45}
  (1980) 1908.

\bibitem{IP} A. Ilakovac and A. Pilaftsis, Nucl.\ Phys.\ B {\bf 437}
  (1995) 491.

\bibitem{LFVrev} A. Pilaftsis, Mod.\ Phys.\ Lett.\ A~{\bf 9} (1994) 3595;\\ 
  M.~C. Gonzalez-Garcia and J.~W.~F. Valle, Mod.\ Phys.\ Lett.\ A~{\bf
  7} (1992) 477; Erratum~{\bf 9} (1994) 2569;\\
  L.~N. Chang, D.  Ng and J.~N. Ng, Phys.\ Rev.\ D {\bf
  50} (1994) 4589;\\ 
  G.  Bhattacharya, P. Kalyniak and I. Mello, Phys.\
  Rev.\ D {\bf 51} (1995) 3569;\\ 
  D.~Tommasini, G.~Barenboim, J.~Bernabeu and C.~Jarlskog, Nucl.\ Phys.\
  B {\bf 444} (1995) 451;\\ 
  A. Ilakovac, B.~A.  Kniehl, and A. Pilaftsis, Phys.\ Rev.\ 
  D {\bf 52} (1995) 3993;\\ 
  A. Ilakovac, Phys.\ Rev.\ D {\bf 54} (1996) 5653;\\ 
  S.  Fajfer and A.  Ilakovac, Phys.\ Rev.\ D {\bf 57} (1998) 4219;\\ 
  M.  Raidal and A.  Santamaria, Phys.\ Lett.\ B {\bf 421} (1998) 250;\\ 
  M.~Czakon, M.~Zralek and J.~Gluza, Nucl.\ Phys.\ B {\bf 573} (2000) 57;\\ 
  J.~I.~Illana and T.~Riemann, Phys.\ Rev.\ D {\bf 63} (2001) 053004;\\ 
  G.~Cvetic, C.~Dib, C.~S.~Kim and J.~D.~Kim, Phys.\ Rev.\ D {\bf 66} (2002)
  034008; 
  Phys.\ Rev.\  D {\bf 71} (2005) 113013;\\ 
  A.~Masiero, S.~K.~Vempati and O.~Vives, New J.\ Phys.\  {\bf 6}
  (2004) 202;\\
  F.~Deppisch and J.~W.~F.~Valle,
  %``Enhanced lepton flavour violation in the supersymmetric inverse seesaw
  %model,''
  Phys.\ Rev.\  D {\bf 72} (2005) 036001; Nucl.\ Phys.\  B {\bf 752} (2006) 80.


\bibitem{KPS} J.~G. K\"orner, A. Pilaftsis and K. Schilcher, Phys.\
  Lett.\ B {\bf 300} (1993) 381;\\
  J.~Bernab\'eu, J.~G.~K\"orner, A.~Pilaftsis and K.~Schilcher, Phys.\
  Rev.\ Lett.\ {\bf 71} (1993) 2695. 

\bibitem{Babar}
  B.~Aubert {\it et al.}  [BABAR Collaboration],
  %``Search for lepton flavor violation in the decay $\tau \to \mu \gamma$,''
  Phys.\ Rev.\ Lett.\  {\bf 95} (2005) 041802.

\bibitem{AJexp} A. Jodidio {\it et al.}, Phys.\ Rev.\ D~{\bf 34}
  (1986) 1967.

\bibitem{ARGUS} ARGUS Collaboration (H. Albrecht {\it et al.}), 
  Z. Phys.\ C~{\bf 68} (1995) 25.

\bibitem{GGR} G.~G.~Raffelt,
  %``Astrophysical methods to constrain axions and other novel particle
  %phenomena,''
  Phys.\ Rept.\  {\bf 198} (1990) 1.

\bibitem{Astro} D.~S.~P.~Dearborn, D.~N.~Schramm and G.~Steigman,
  %``Astrophysical constraints on the couplings of axions, majorons, and
  %familons,''
  Phys.\ Rev.\ Lett.\  {\bf 56} (1986) 26;\\
 H.~Y.~Cheng,
  %``TABULATION OF ASTROPHYSICAL CONSTRAINTS ON AXIONS AND NAMBU-GOLDSTONE
  %BOSONS,''
  Phys.\ Rev.\  D {\bf 36} (1987) 1649;\\
R.~Chanda, J.~F.~Nieves and P.~B.~Pal,
  %``ASTROPHYSICAL CONSTRAINTS ON AXION AND MAJORON COUPLINGS,''
  Phys.\ Rev.\  D {\bf 37} (1988) 2714;\\
K.~Choi and A.~Santamaria,
  %``MAJORONS AND SUPERNOVA COOLING,''
  Phys.\ Rev.\  D {\bf 42} (1990) 293.

\bibitem{NI} N.~Iwamoto, %``Axion Emission From Neutron Stars,''
  Phys.\ Rev.\ Lett.\ {\bf 53} (1984) 1198.

\bibitem{BKLMS} 
V.~Barger, J.~P.~Kneller, H.~S.~Lee, D.~Marfatia and G.~Steigman,
  %``Effective number of neutrinos and baryon asymmetry from BBN and WMAP,''
  Phys.\ Lett.\  B {\bf 566} (2003) 8;\\
S.~Hannestad,
%``Neutrino masses and the number of neutrino species from WMAP and  2dFGRS,''
  JCAP {\bf 0305} (2003) 004.

\bibitem{IST} For a recent study, see,\\
K.~Ichikawa, T.~Sekiguchi and T.~Takahashi,
%``Probing the Effective Number of Neutrino Species with Cosmic
%Microwave Background,''
arXiv:0803.0889 [astro-ph].


\bibitem{ASJ} A.~S.~Joshipura and J.~W.~F. Valle, Nucl.\ Phys.\ 
B~{\bf 397} (1993) 105;\\
F.~De Campos, M.~A.~Garcia-Jareno, A.~S.~Joshipura, J.~Rosiek,
J.~W.~F.~Valle and D.~P.~Roy, 
%``Limits on associated production of visibly and invisibly decaying Higgs
%bosons from Z decays,''
Phys.\ Lett.\  B {\bf 336} (1994) 446;\\
D.~G.~Cerdeno, A.~Dedes and T.~E.~J.~Underwood,
%``The minimal phantom sector of the standard model: Higgs phenomenology  and
%Dirac leptogenesis,''
JHEP {\bf 0609} (2006) 067.



\end{thebibliography}
\end{document}